\pdfoutput=1  

\def\prePrint{1}            
\def\generateTikz{1}        

\if\prePrint0
    \documentclass[subscriptcorrection,upint,varvw,barcolor=Goldenrod3,mathalfa=cal=euler,balance,hyphenate,french,pdf-a]{asmejour} %

    \hypersetup{%
        pdfauthor={Jonathan T. Horne-Jones},                       		   	
        pdftitle={Towards Virtual Qualification in Fusion: },                  	
        pdfkeywords={Fusion, Virtual Qualification, Validation, Simulation, Probabilistic},
        pdfsubject = {Paper describing work on credible model validation in a fusion context, working towards realisation of virtual qualification in fusion},			
    }

    \JourName{Journal of Verification, Validation and Uncertainty Quantification}

\else
    \documentclass[a4paper,9pt,twocolumn]{extarticle}

    \RequirePackage[a4paper, margin=2cm]{geometry}
    \RequirePackage[sort&compress,numbers]{natbib}
    \RequirePackage{graphicx}
    \RequirePackage{array}
    \RequirePackage{booktabs}
    \RequirePackage{mathtools}
    \RequirePackage[]{babel}
    \RequirePackage{bm}
    \RequirePackage[]{newtxtext} 
    \RequirePackage[varl,varqu]{inconsolata}
    \RequirePackage[]{newtxmath}

    \addto{\captionsenglish}{

    }

    \RequirePackage[labelfont={sf,bf},hypcap=false]{caption}
    \RequirePackage[hypcap=false,list=true]{subcaption}
    \RequirePackage{authblk}
        \RequirePackage{hyperref}
    \RequirePackage[hyperref,dvipsnames,svgnames,x11names]{xcolor}

\fi

\usepackage{tikz}
\usepackage{pgfplots}
\usepackage{pgfmath}
\usepackage{tikzscale}

\usetikzlibrary{fit,positioning,tikzmark}
\usetikzlibrary{decorations.markings}
\usetikzlibrary{plotmarks}
\usetikzlibrary{external}

\if\generateTikz1
    \newcommand{\inputtikz}[1]{%
        \tikzsetnextfilename{#1}%
        \input{#1.tikz}%
    }
\else
    \newcommand{\inputtikz}[1]{%
        \includegraphics{figures/#1.pdf}%
    }
\fi

\usepackage{soul}
\usepackage{xcolor}
\usepackage{caption}
\usepackage{subcaption}
\usepackage{wasysym}
\usepackage{textcomp}
\usepackage{placeins}

\begin{document}

\if\prePrint0
    \SetAuthorBlock{Jonathan T. Horne-Jones \CorrespondingAuthor, Michelle Baxter, Adel Tayeb, Lloyd Fletcher, James Paterson, Stephen Biggs-Fox, Allan Harte}{
    United Kingdom Atomic Energy Authority,\\
    Culham Campus,\\
    Abingdon, OX14 3DB, UK \\
    email: Jonathan.Horne-Jones@ukaea.uk}
\else
    \author[1]{J. T. Horne-Jones}
    \author[2]{M. Baxter \thanks{Corresponding author email: michelle.baxter@ukaea.uk}}
    \author[2]{A. Tayeb}
    \author[2]{L. Fletcher}
    \author[1]{J. Paterson}
    \author[2]{S. Biggs-Fox}
    \author[2]{A. Harte}
    \affil[1]{UK Atomic Energy Authority, Culham Campus, Abingdon, OX14 3DB, UK}
    \affil[2]{UK Atomic Energy Authority, Unit 2a Lanchester Way, Rotherham, S60 5FX, UK}
\fi

\title{Towards virtual qualification in nuclear fusion: demonstrating probabilistic model validation on a high heat flux component}

\if\prePrint0
    \keywords{Fusion, Data-rich Experiments, Simulation, Probabilistic, Uncertainty Quantification, Validation, Virtual Qualification}

    \date{Version \versionno, \today}
\else
    \date{}
\fi

\maketitle 

\begin{abstract}
Qualification of components operating in future fusion power plants will be heavily reliant on simulations of component behaviour.
The lack of representative test environments for many aspects of the expected operating environment will necessitate full or partial \emph{virtual qualification} of components.
The cornerstone of virtual qualification is credible validation of the simulation models on which it relies.
In this work, we present a probabilistic model validation framework that forms the basis for implementation of virtual qualification in fusion.
We demonstrate our framework on a representative component; a high heat flux heat sink subject to a tightly coupled multi-physics loading.
We perform data-rich, optimised experiments, in which we implement high fidelity diagnostics and rigorously quantify aleatoric and epistemic uncertainty on all measurements.
Our simulation approach efficiently samples input uncertainty distributions to predict probability boxes describing component response, using a statistical surrogate to replicate behaviour of the finite element model we wish to validate.
We then used a novel implementation of the modified area validation metric to quantify the model form error of the finite element model, isolating it from the aleatoric and epistemic experimental uncertainty.
We discuss the contribution of our validation approach towards virtual qualification, and the benefits of the risk-based decision-making it facilitates.
The experimental, simulation, and validation datasets are published as a benchmark of a probabilistic validation approach for fusion, and for use in development of new model validation methodologies.
\end{abstract}

\if\prePrint1
    {\bf Keywords:} Fusion, Data-rich Experiments, Simulation, Probabilistic, Uncertainty Quantification, Validation, Virtual Qualification
\fi


\section{Introduction} \label{sec:Introduction}

\subsection{Fusion energy}
Nuclear fusion is the process of combining small nuclei to form a larger nucleus, and in doing so releasing energy.
Achieving fusion requires high temperatures and pressures, conditions in which the hydrogen fuel ionises and forms a plasma.
Key to sustaining a fusion plasma is the ability to confine it, maintaining the conditions required for fusion while also protecting the surrounding components \citep{stacey2010fusion}.
Fusion energy research has primarily focussed on two confinement approaches: magnetic confinement fusion \cite{stacey2010fusion,ongena2016magnetic} and inertial confinement fusion \citep{betti2016inertial}.
In this work we focus on magnetic confinement fusion, though the methods we will present are applicable to both.
In magnetic confinement fusion, the plasma is contained within a vacuum vessel, and is shaped and controlled by multiple sets of large magnets in a device called a tokamak \citep{wesson2011tokamaks}.
Components located within the vacuum vessel (called in-vessel components) are designed to absorb heat from the plasma, both to harness it for energy production and to protect the other systems in the tokamak.
These components are subject to the most extreme conditions of any system within a tokamak, and their performance and reliability are critical to overall plant operation.

The concept of harnessing fusion for energy generation has long been a topic of scientific research, but in recent years there have been increased efforts towards realisation of a commercially viable fusion energy source \cite{chapman2024spherical}.
The first generation of prototypic fusion power plants represent a step change in the loading conditions for in-vessel components \citep{taylor2014lessons,horvath2016nuclear}.
With the aim to move to higher powered continuous operation, the loading the components will be subjected to will significantly increase, most notably the heat and neutronics loads \citep{wenninger2017demo,stork2014materials}.
Prior to operation, there will be no mechanism by which to physically test in-vessel components under the exact environmental conditions of a fusion power plant \citep{gorley2018materials}.
This presents the challenging problem of how to qualify fusion components to the satisfaction of key stakeholders, such as investors or a safety regulator, using a simulation-led approach.
If components cannot be qualified, commercial fusion will not be realised.

\subsection{Component qualification in fusion} \label{subsec:QualificationFusion}

Much of the past and current focus in fusion component qualification is on development and usage of design codes and standards \citep{sannazzaro2013development,davis2023need}.
It is well documented that existing design codes are inadequate for the unique demands on components in future fusion power stations \citep{gorley2018materials,aiello2011assessment,davis2023need,porton2016structural}.
A key limitation is that they are deterministic, and rely on conservatism to demonstrate confidence in component performance and safety.
Each of the numerous sources of loading on fusion components has associated uncertainty that must be accounted for in deterministic design codes by increased conservatism.
Widespread design conservatism is incompatible with the need for fusion to be commercial viable in a competitive energy market, and, as such, new approaches to fusion design criteria and qualification must be considered.
Gorley et al. \citep{gorley2018materials} discuss a number of potential approaches, including the use of probabilistic analysis to quantity variability of component performance due to input uncertainty.
Not only does this facilitate risk-based assessment of component performance, it is also more tolerant to sparsity of physical test data than traditional deterministic methods.
With the aforementioned lack of physical test data for future fusion environments \cite{gorley2018materials}, this takes on increased significance.

Lawless et al. \cite{lawlessoverview} present an overview of the implementation of a ``digital first" approach to fusion system design and qualification at the UK Atomic Energy Authority (UKAEA).
This represents a move away from historical reliance on physical testing, and instead relies on extensive use of simulation to accelerate the development path towards commercial fusion.
Central to this approach is the concept of \emph{virtual qualification}, defined as the process of using simulations to determine the probability that a component will meet its functional and system requirements.
This encapsulates the improvements to traditional qualification discussed above, in that it is simulation based and computes output probability.
We envisage a framework for virtual qualification incorporating the roadmap that Lawless et al. \cite{lawlessoverview} describe, with the following features.
\begin{enumerate}
    \item Simulation validation is performed where test facilities are available. A heavy emphasis is placed on maximising value from any experiments by using high fidelity, data-rich diagnostic methods.
    \item Validation of directly measured model outputs is interpolated to unmeasured responses of interest.
    \item Separately validated simulation models are combined to provide predictive capability of component performance, including aggregation of model form uncertainty.
    \item Where test facilities do not sufficiently represent power plant operating conditions, model validation is assimilated to those operating conditions using understanding of the basis physics.
    \item Decisions makers are provided with probabilistic qualification evidence that allows risk-based decision-making and informed assignment of budget to improving component and/or model performance.
\end{enumerate}

Application of virtual qualification to real-world fusion components is the ultimate aim of the work we present here.

\subsection{Probabilistic model validation} \label{subsec:IntroModelValidation}

A core feature of a virtual qualification framework, and an unavoidable requirement for use of simulation in any qualification, is credible model validation.
It is well understood that models are always an approximation to real-world behaviour \cite{box1976science}.
The role of model validation is to demonstrate that the assumptions inherent to the modelling method are an acceptable approximation, given the context of use of the model.
The methods and requirements for model Verification Validation and Uncertainty Quantification (VVUQ) have been documented by a range of organisations \cite{nafems2020esqms,nasa2024standard,asme2019vv10,asme2018vv40,oberkampf2007predictive}, in all cases to ensure the resultant validation is credible.
While there is some variation between these standards, a common theme is the importance of robust characterisation of uncertainty in any validation experiment and in the associated model.
Implementation of these standards therefore requires a probabilistic approach to experiments, simulation, and model validation.

\begin{figure*}
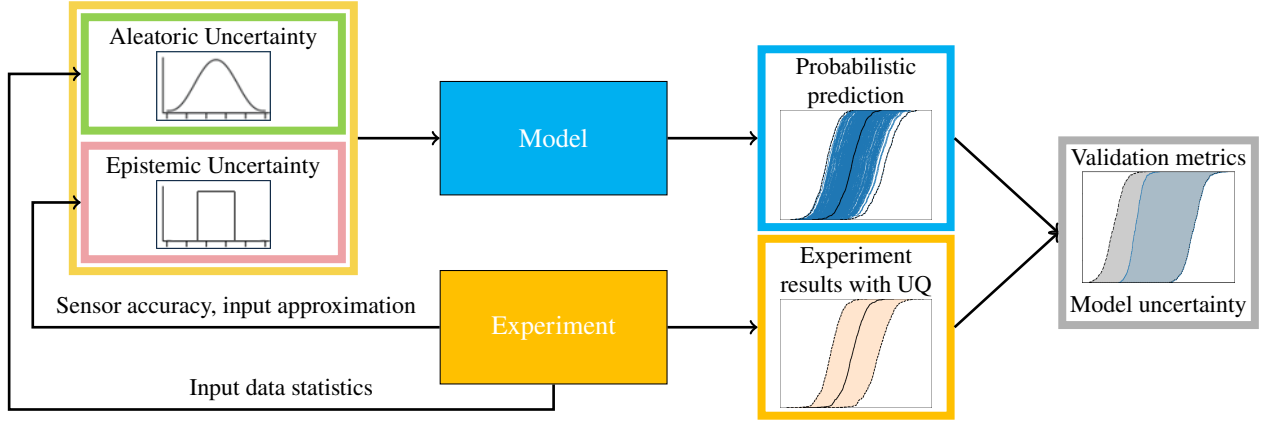

    \centering
    \inputtikz{validation_workflow}
    \caption{A probabilistic validation workflow}
    \label{fig:validation_workflow}
\end{figure*}

In all real world systems, and therefore also in models of those systems, there exists two types of uncertainty.
\emph{Aleatoric} uncertainty is irreducible uncertainty due to inherent randomness, for example noise on a measurement.
\emph{Epistemic} uncertainty is that due to lack of knowledge, for example some systematic error in an experimental setup.
Epistemic uncertainty is the more challenging to process, because the lack of knowledge requires its representation as an interval about a nominal value.
Figure~\ref{fig:validation_workflow} shows a workflow for validation of a model that is representative of the approach we will follow in this paper.
The features discussed above are evident.
The aleatoric and epistemic uncertainties computed at experiment input facilitate probabilistic implementation of the simulation model, and uncertainties computed at output provide probabilistic experimental results against which to evaluate model form uncertainty.

\subsection{Overview}
In this paper, we present the foundation to a future virtual qualification framework for fusion components.
We focus on credible model validation, and present an application of VVUQ to a model of a representative fusion component and operating scenario.
We aim to satisfy the rigorous VVUQ requirements set out in standards such as ASME V\&V-40 \citep{asme2018vv40} and demonstrated by McVeigh et al. \citep{mcveigh2023perspectives}, and therefore present our approach and results to a high level of detail.
We first discuss selection of a suitable initial demonstrator in \S\ref{sec:ModelValidationDemo}.
This is followed by incorporation of the demonstrator component into optimised, data-rich validation experiments with rigorous uncertainty qualification in \S\ref{sec:Experiments}.
We present our probabilistic simulation approach in \S\ref{sec:Simulations}, in which we detail our approach to multi-physics finite element modelling, surrogate modelling, sensitivity analysis, input sampling and uncertainty propagation.
We validate our simulation results using a novel validation metric implementation in \S\ref{sec:Validation}, accounting for epistemic and aleatoric uncertainty in the both the simulation and experimental domains and validating full field component response.
The performance of the methodology, the benefits it will give to fusion, and plans for follow on work are discussed in \S\ref{sec:Discussion}.


\section{A model validation demonstrator for fusion} \label{sec:ModelValidationDemo}

In order to develop a model validation methodology suitable for the virtual qualification framework, we required a fusion relevant test case.
Moreover, we required an initially simple test case that can be evolved in future work to incorporate the various degrees of complexity of a realistic fusion operating environment. 
One of the most challenging areas of fusion component design is achieving adequate performance of plasma facing components (PFCs) \cite{cane2024managing}.
These components are subjected to complex multi-physics loading, combining high surface heat flux, electromagnetic loads and heating, neutronic heating and material damage, and plasma-surface interaction (e.g. melting, vaporising, sputtering etc.) \citep{barrett2016progress,cane2024managing,maviglia2018wall,maviglia2020impact,reinhart2022latest,you2022divertor}.
Their steady state loading is generally dominated by the surface heat flux from the adjacent plasma \citep{barrett2016progress}, which motivates our choice to initially develop our model validation methodology on surface heating of a PFC.
Additional complexity is readily available for future work, in which we will seek to incorporate modelling and validation of electromagnetic and neutronic loading.
This also aligns well with the availability of test facilities within UKAEA.
Small-scale, high surface heat flux testing is already facilitated by the HIVE facility \citep{pearl_cyclic_2023}, soon to be followed by the LIBRTI \citep{lawlessoverview} (neutron source) and CHIMERA \citep{barrett2023chimera} (combined heating and magnetic loading) facilities.

\begin{figure}
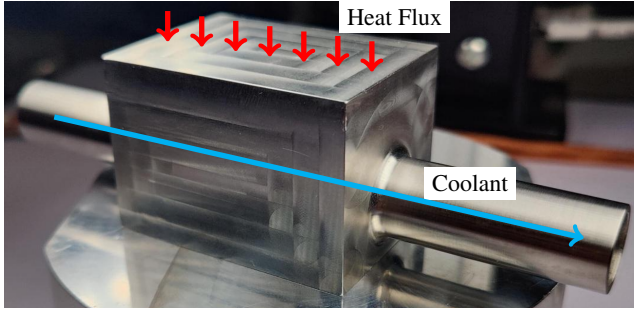

    \centering
    \inputtikz{monoblock_image}
    \caption{The single material, 316L stainless steel, monoblock used within this work.}
    \label{fig:monoblock_image}
\end{figure}

In this work we use a monoblock, a candidate for very high heat flux heat sink components in fusion \citep{hirai2015status}, as our component under test.
Conventionally, monoblocks have a multi-material construction, with a tungsten plasma facing armour, copper chrome zirconium coolant pipe, and copper interlayer connecting the two with brazed or hot radially pressed joints \citep{gavila2011high}.
The multi-material system with joining introduces significant complexity to the system for this initial methodology development work (e.g. residual stress from manufacturing), and we therefore amend the monoblock design to remove the requirement for any bonding and allow machining from a single material.
This satisfies our requirement for a simple, but relevant and extensible, initial test case.
The model validation demonstrator is shown in Fig.~\ref{fig:monoblock_image}.

\section{Data-rich experiments} \label{sec:Experiments}

We performed experiments to study the thermal and mechanical response of the monoblock component introduced in \S\ref{sec:ModelValidationDemo} to surface heat flux, akin to loading of this type of component in a fusion reactor.
For this initial demonstration of model validation, we target steady state behaviour within the elastic deformation regime.
Central to our validation approach is implementation of data-rich experiments, in which we capture a large number of spatially distributed data points.
We do so by implementation of an image based diagnostic, specifically Digital Image Correlation (DIC), alongside collections of point sensors.
This provides us with valuable insight into the spatial structure of the component response, and facilitates a high fidelity validation.
The requirement for data-rich experiments drives our choice of experimental facility and diagnostic methods.

\subsection{Experimental setup} \label{subsec:ExperimentalSetup}

\subsubsection{Sample preparation}
The monoblock was machined from a solid block of 316L stainless steel, to the design and dimensions shown in fig \ref{fig:experiment_monoblock}.
This manufacturing approach allowed for no in-component joints and minimised unintentional localised heat treatment of the material, thereby avoiding addition of uncertainty in material properties or the need to account for residual stress due to manufacturing processes.
Small holes, of $2~\mathrm{mm}$ nominal diameter and $0.5~\textrm{mm}$ nominal depth, were machined into the surface of the monoblock to facilitate repeatable and accurate installation of the thermocouples at their designated positions.
The holes are considered sufficiently small so as to minimally affect the response of the monoblock and not require incorporation into the model.

We applied a high contrast speckle pattern for DIC to the front face of the monoblock and induction coil, using an airbrush to apply VHT flameproof paint.
This paint is tolerant to temperatures of up to $1000~^{\circ}C$, well within the temperature response range of the monoblock at steady state.
The fineness of the speckle pattern features contributes to the resolution of the DIC measurements, with our experiments using pattern features corresponding to approximately 8 image pixels and image speckle size of $240~\mu\textrm{m}$.

\subsubsection{Experimental facility: HIVE}
\begin{figure*}
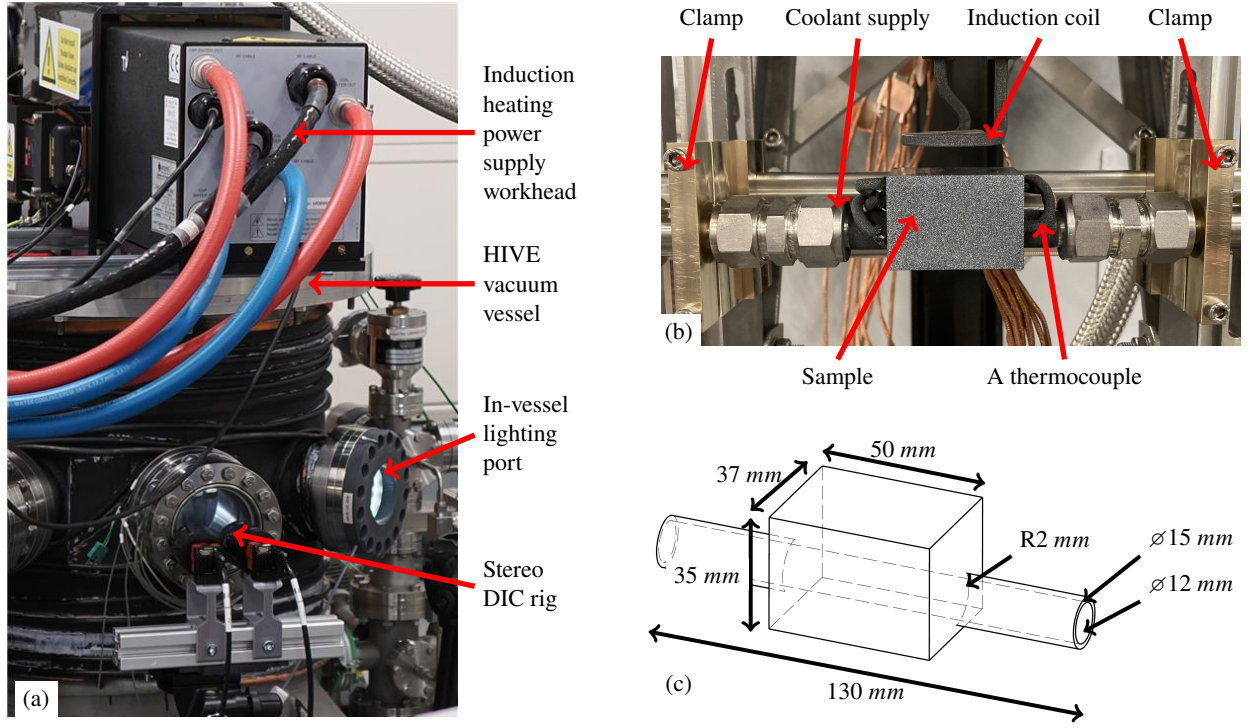

    \centering
    \inputtikz{experiment}
    {\phantomsubcaption\label{fig:experiment_exterior}
     \phantomsubcaption\label{fig:experiment_interior}
     \phantomsubcaption\label{fig:experiment_monoblock}}
    \caption{The experimental setup. The HIVE vessel and induction heating system is shown in (a) along with the locations of ex-vessel diagnostic systems. (b) shows the monoblock and coil in-vessel, both with painted speckle pattern to facilitate DIC measurements. The design and dimensions of the monoblock is shown in (c).}
    \label{fig:experiment}
\end{figure*}

We conducted our validation experiments at UKAEA's aforementioned Heating by Induction to Verify Extremes (HIVE) facility \citep{pearl_cyclic_2023}.
HIVE is a high heat flux testing facility designed to test small components under fusion relevant surface heat fluxes under vacuum.
It uses a high frequency induction heating system, operating at frequencies of $\mathcal{O}(100~\textrm{kHz})$ with a peak power of $45~\textrm{kW}$, utilising the phenomena of electromagnetic skin effects in conductive materials to concentrate induced currents in a very thin surface-adjacent layer and provide a proxy to surface heating.
Samples under test are cooled using a closed-loop water cooling system, operating at up to $T_{cool}=200~^{\circ}\textrm{C}$, $p_{cool}=20~\textrm{bar}$, and $\dot{Q}_{cool}=90~\textrm{L min}^{-1}$.
Experiments are conducted within a stainless steel vacuum vessel, the lid of which supports the sample under test and provides feedthroughs for induction heating supply, coolant, and multiple in-vessel diagnostic channels.
The main body of the vacuum vessel incorporates multiple view ports at the vertical level of the sample under test, through which optical diagnostics and lighting can be used.
The HIVE vessel and heating system can be seen in fig \ref{fig:experiment_exterior} and the integration of our monoblock with the HIVE heating and cooling system in fig \ref{fig:experiment_interior}.
For further information on HIVE, the reader is referred to Pearl et al.  \citep{pearl_cyclic_2023}.

\subsubsection{Diagnostics} \label{subsubsec:Diagnostics}
We utilised the diagnostic flexibility of HIVE to combine on-sample measurements of temperature with image-based measurement of the mechanical response using stereo DIC with ex-vessel cameras.
We recorded the surface temperature across the monoblock using a set of 10 K-type thermocouples.
Additional thermocouples were used to measure coil and monoblock coolant temperatures up- and downstream of the respective components.
Measurements of the coolant pressure and flow rate were recorded using Omega PX419-750GI-EH pressure sensors, and Nixon NT13 flowmeter, respectively.
The above diagnostics were sampled at a rate of $1~\textrm{Hz}$ using two digital acquisition systems (DAQs): a National Instruments cDAQ-9174 linked to the DIC data stream, and a National Instruments cRIO-904 linked to the HIVE system.
Both systems facilitated cold-junction temperature compensation for all thermocouples used.
The HIVE DAQ was configured to output a measure of heating power level to an analogue data channel that was taken as input to the DIC DAQ in order to synchronise the data captured by the two separate systems.

A key input to the simulations of the experiment is the current through the induction coil. We used a PEM CWT30xB Rogowski coil, mounted around one of the induction heater feedthroughs, to measure this current.
Additional data on coil behaviour was recorded using a Rigol RP1100D high voltage differential probe.
Both data streams were captured with a PicoScope 3406D MSO, taking an average of the root mean square values of the high frequency signals over at least 0.1s to remove fluctuations from rectification and active control.

We took image-based measurements of the mechanical response of the front face of the monoblock using stereo DIC.
This technique captures images of the face of the monoblock, recording the movement of a high contrast speckle pattern on the surface.
These images are then post-processed to infer and track the locations of a set of subsets of pixels on the surface throughout an experiment.
We used a stereo DIC setup, positioned a small, known lateral distance apart giving a stereo angle of $11.2^\circ$.
Once calibrated, this allows measurement of out of plane displacement of the surface of the monoblock which is essential due to volumetric expansion of the sample.
In our DIC setup we used two 24.6 MPx Alvium 1800 U-2460 cameras, each with a 35 mm fixed focal length lens.
We implement our DIC setup in accordance with the International Digital Image Correlation Society best practice guidance \cite{jones2025dicpractice}, and report calibration details in Appendix \ref{ap:dicCalibration}.
We assess the resolution of our diagnostic by computing the temporal standard deviation of results over a set of 100 images captured before the experiments to compute the noise floor.
We find a maximum temporal noise floor in the components of strain of interest of $116.8~\mu\epsilon$ (full details given in Table \ref{tab:NoiseFloor} in Appendix \ref{ap:dicCalibration}).

\subsubsection{Experimental protocol}
We ran heating pulses with an induction system power of $21.5~\textrm{kW}$ using a bespoke induction coil excited at a frequency of $123~\textrm{kHz}$.
Coolant water was applied through the components pipe at $T_{cool}=150~^{\circ}C$, $p_{cool}=6~\textrm{bar}$, and $\dot{Q}_{cool}=50~\textrm{L min}^{-1}$ to achieve response with significant temperature variation, but within the elastic regime.
To access steady state behaviour, we performed experiments with a total duration of $1150~\textrm{s}$.
This included an initial $100~\textrm{s}$ with coolant flow but no heating, while recording all sensor traces, which provides a well understood initial condition and baseline noise floor for our sensors.
Heating was applied for a total duration of $700~\textrm{s}$, of which the final $\sim 400~\textrm{s}$ is considered steady state thermal response.
The final $350~\textrm{s}$ maintains coolant flow without heating and is of sufficient duration for the monoblock to return to the initial condition. 
We ran three repeated pulses of the above conditions using a single monoblock sample.

\subsection{Experiment optimisation} \label{subsec:ExperimentOptimisation}

We used a pre-test simulation to drive optimisation of our experimental setup, with two particular focuses: placement of point sensors; and optimal monoblock sizing and placement.
This used the same simulation method as is presented in \S\ref{subsec:FEA}, but with initial estimates of input values used as no experiment-derived values were available.
These are described in \S\ref{subsubsec:SensorOptimisation} and \S\ref{subsubsec:SampleSizePlace} respectively.

\subsubsection{Sensor placement} \label{subsubsec:SensorOptimisation}

Validation of a model using experimental data is reliant on the data from the experiment sufficiently capturing the behaviour of the component under test.
A common limitation, and one that is likely in operational fusion components \citep{raukema2024demonstration}, is sparsity of data.
In our experiments, the data on the temperature field is limited to that from 10 thermocouples, themselves limited to quasi-surface placement on the back and side faces of the monoblock.
This is to avoid the camera view face and to position the thermocouples away from the induction coil.

\begin{figure}
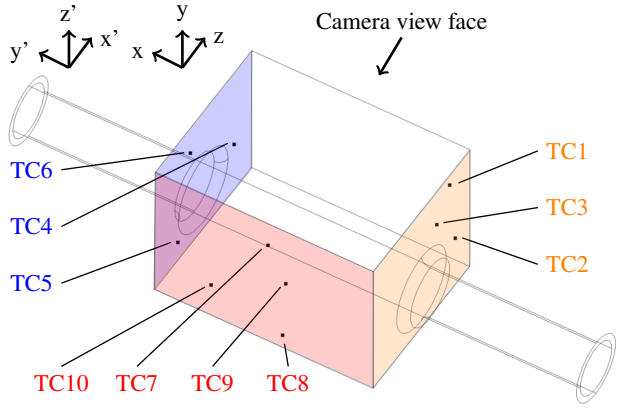

    \centering
    \inputtikz{thermocouple_placement}
    \caption{The optimised thermocouple locations on the monoblock. Two coodinate systems are shown: the DIC image system (x,y,z), and the simulation model system (x',y',z').}
    \label{fig:thermocouple_placement}
\end{figure}

To maximise the value of the data captured by these thermocouples, we employed an in-house optimisation tool to place the thermocouples within the permissible surface areas on the monoblock.
The optimisation uses temperature data from the pre-test simulations to characterise the behaviour of the monoblock.
At each trial sensor location, we extract simulated thermocouple readings and train a Gaussian Process model to reconstruct the temperature field from these readings.
The cost function is the error between the full 3D simulation data and the reconstructed temperature field from the Gaussian process model summed over the volume.
The optimisation was constrained to maintain a $4~\textrm{mm}$ spacing between thermocouples to allow for their installation, and similarly to position the thermocouples at least $2~\textrm{mm}$ from the surface edges and coolant pipe.
The trial locations were generated using Bayesian optimisation, utilising the Bayesian Optimization python library \citep{baysopt}.
We perform this in two passes, the first generating 100 locations using an expected improvement acquisition function (5 of which are an initial random seeding).
In the second pass we use a probability of improvement acquisition function to generate an additional 100 trial locations.
The optimised temperature field reconstruction achieves a maximum error to simulation of $30~^\circ \textrm{C}$, in the un-instrumented upper top surface of the monoblock.
The resultant thermocouple locations are shown in fig \ref{fig:thermocouple_placement}.
Coordinates (as used in the simulation model) are given in Table \ref{tab:ThermocouplePlacement} in Appendix \ref{ap:ThermocouplePlacement}.

\subsubsection{Monoblock sizing and placement} \label{subsubsec:SampleSizePlace}
Minimisation of uncertainty is an important aspect of design of validation experiments.
Much of this is achieved by selection of appropriate diagnostics and good understanding of the as-tested experimental setup.
Key to the above is understanding the sensitivity of the system to the various aspects of the experiment design.
Uncertainty in experiment-derived system inputs is exacerbated if the system is particularly sensitive to those inputs.
A validation experiment can therefore be optimised by amending the design to reduce the sensitivity on an input for which the experiment unavoidably introduces uncertainty.

We utilised pre-test simulations to run an initial sensitivity study for use in experiment design.
We identified relative induction coil to monoblock translation and rotation as the dominant inputs, particularly the vertical distance between the two.
High sensitivity in these inputs is problematic, as coolant flow in both objects drives vibrations with cumulative amplitude of up to $0.5~\textrm{mm}$.
The induction coil and monoblock also magnetically repel and therefore deflect slightly when the coil is energised.
Typical HIVE experiments run with a coil-sample gap of $2~\textrm{mm}$, for which the signal-to-uncertainty ratio is poor.
Additionally, the heating of the monoblock peaks at the front and rear edges, where the induced heating currents concentrate.
This results in sensitivity to the lateral relative translation and rotation.

We amended the design to mitigate the above issues.
We first significantly increased nominal distance between the induction coil and monoblock to give a gap of $10~\textrm{mm}$, increasing the coil current to maintain the same heating power in the monoblock.
We also increased the size of the monoblock, such that the upper surface more than spanned the whole induction coil.
These resulted in significant reduction in response sensitivity to the known geometric uncertainties of the experiment.

\subsection{Experiment post-processing} \label{subsec:ExperimentPostprocessing}
All experiments require some degree of post-processing to extract useful information from the raw data.
In validation experiments, additional weight is placed on the \emph{credibility} of any post-processing performed, as it must not bias the experimental results in any way, to allow for accurate quantitative validation of simulation models.
We have performed post-processing of our thermocouple and DIC data, ensuring to quantify different sources of uncertainty from the experiment and the post-processing itself.
This allows us to account for this uncertainty when performing our validation analysis.

\subsubsection{Thermocouple correction and selection} \label{subsubsec:Thermocouples} 
We used unshielded, surface mounted thermocouples to reduce uncertainty in thermal contact location and aid in installation weld quality.
However, the hostile operating environment in HIVE and handling of the monoblock during installation resulted in faults and artefacts in the data, the resolution of which is discussed below.

The first category of problems with the thermocouples is some type of fault with the sensor and/or installation.
We found that thermocouples TC1 and TC7 generated no data, with the case for omission from the study therefore clear.
TC4 did generate data, but we observed significant discrepancy between its readings and those of all other thermocouples, including those directly measuring coolant temperature, before the heating commences in all experiment repeats.
Measurements are taken for $100~\textrm{s}$ before the start of heating, with the coolant flowing, during which the monoblock will be at coolant temperature.
As the discrepancy in TC4 results during this period was greater than the quoted accuracy of the thermocouples, we assert that the data is invalid and should be omitted.

\begin{figure}
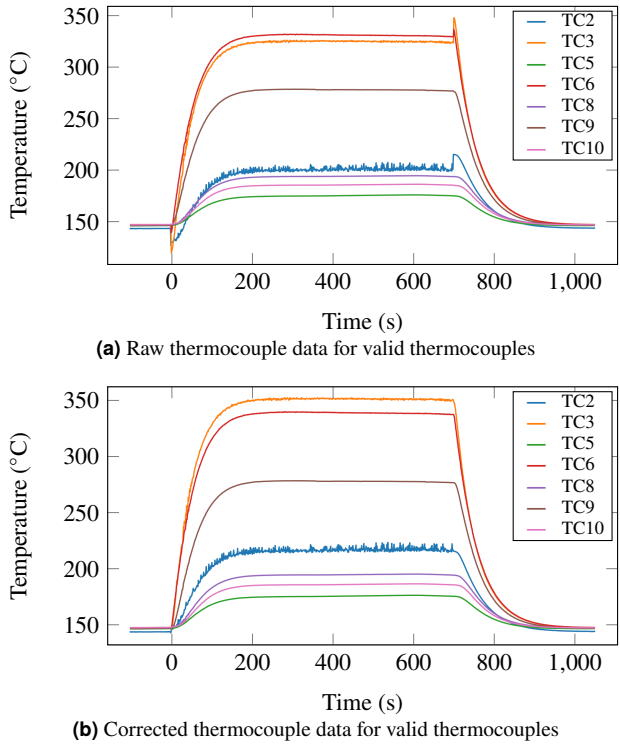

    \centering
    \begin{subfigure}{\linewidth}
        \centering
        \inputtikz{thermocouples_raw}
        \vspace{-0.25cm}
        \caption{Raw thermocouple data for valid thermocouples}
        \label{fig:thermocouples_raw}
    \end{subfigure}
    \begin{subfigure}{\linewidth}
        \vspace{0.25cm}
        \centering
        \inputtikz{thermocouples_corrected}
        \vspace{-0.25cm}
        \caption{Corrected thermocouple data for valid thermocouples}
        \label{fig:thermocouples_corrected}
    \end{subfigure}
    \caption{Temperature traces for all valid thermocouples through one of the experiments, pre- (a) and post- (b) correction for interference from the induction coil. The raw data show step changes in reported temperature at the start and end of induction coil operation that is removed in the corrected data.}
    \label{fig:thermocouple_data}
\end{figure}

When placing our thermocouples, we avoided the top $5~\textrm{mm}$ of the sample in an attempt to minimise interference due to the proximity of the high frequency induction coil.
When combined with the usage of unshielded thermocouples, this proved to be inadequate and resulted in the interference on some thermocouple data shown in fig \ref{fig:thermocouples_raw}.
Cross-reference of induction coil current data with the period in which an offset is visible in the reported temperature confirmed that the induction coil was responsible.
The size of the offset would be expected to scale with the magnitude of coil current and proximity to the coil, but incidental shielding of some thermocouple wires resulted in unpredictable variability between different thermocouples.
To correct for this, we apply the inverse of the average of the offsets measured at start and end of pulse
\begin{equation} \label{eq:thermocouple_correction}
    T_{final} = T_{raw} - \frac{\Delta T_{start} + \Delta T_{end}}{2},
\end{equation}
where $|\Delta T|$ denotes the magnitude of the maximum offset measured between consecutive data points at start and end of heating.
The magnitude of the offset was seen to vary in time, a behaviour (\ref{eq:thermocouple_correction}) does not capture. We therefore account for the discrepancy in offset as an addition to the epistemic uncertainty budget of each thermocouple data stream of $\pm 0.5(|\Delta T_{start}| - |\Delta T_{end}|)$.
The thermocouple temperature traces for one of the experiments, post-correction, are shown in fig \ref{fig:thermocouples_corrected}.

The correction methodology described above provides an indicative example of an approach that will be required in future, large scale, fusion component testing.
Facilities like the aforementioned CHIMERA and LIBRTI rigs will perform testing in much more extreme environments than the one we have presented here.
Component qualification experiments in these facilities will be expensive, and it will rarely be acceptable to re-run experiments in the event of corrupted or biased sensor data.
We will require robust and systematic methods to discard sensors or add additional uncertainty to enable retention of valuable data.
Our approach to this should be used as a baseline from which to develop these methods.

\subsubsection{Post-processing DIC data} \label{subsubsec:DICPostProcessing}
To obtain strain fields on the surface of the monoblock, a two step DIC process is followed.
First, we determine the displacement fields on the monoblock surface by minimising the grey level mismatch over all square-shaped pixel subsets of the image over time.
We then computed the strain fields from these displacement components by differentiation with respect to the two in-plane coordinates.
We used MatchID software \citep{MatchID} for both DIC and strain calculations.
After obtaining the displacement fields on the surface, MatchID fits an analytical approximation of the displacement fields as either bilinear (Q4) or bi-quadratic (Q8) polynomial over a square strain window containing $N\times N$ displacement data points (i.e. pixel subsets).

To identify the optimal DIC post-processing parameters, we performed an optimisation process over $\mathcal{O}(10^3)$ DIC parameter combinations to identify the combination that best minimises the strain fields between the FE results and DIC-levelled data.
The DIC-levelled strains are obtained by numerically deforming experimental images based on FE results from the preliminary simulations and performing both displacement and strains calculations with them.
This process was first introduced by Lava et. al \citep{lava2009} and applied to a similar component to our test sample in the HIVE facility by Tayeb et. al \citep{tayeb2025image-based}.
We found that the optimal DIC parameters are 27 pixels for the subset size and strain window with 27 data points with a bilinear displacement approximation polynomial.

\subsubsection{As-tested geometry} \label{subsubsec:AsTestedGeometry}

We noted in \S \ref{subsubsec:SampleSizePlace} that the response of the monoblock to the heat load is highly sensitive to the as-tested relative position and orientation of the induction coil and monoblock.
While we have amended the experiment design to reduce this sensitivity, these inputs are still the most influential inputs to the simulation.
We therefore required measurement of the as-tested relative position of the coil and sample, and used the DIC data to achieve this.

The front faces of the monoblock and induction heating coil were both given a speckle pattern.
The DIC analysis then generates the locations of a set of coordinates describing each surface in 3D space, which represents the as-tested geometry.
The simulation model is generated from the nominal, or as-designed, geometry, and in order to best model the experiment and understand the geometric input uncertainty we use a post-processing step to determine the transformation required to map the as-designed geometry onto the as-tested geometry.

\begin{figure}
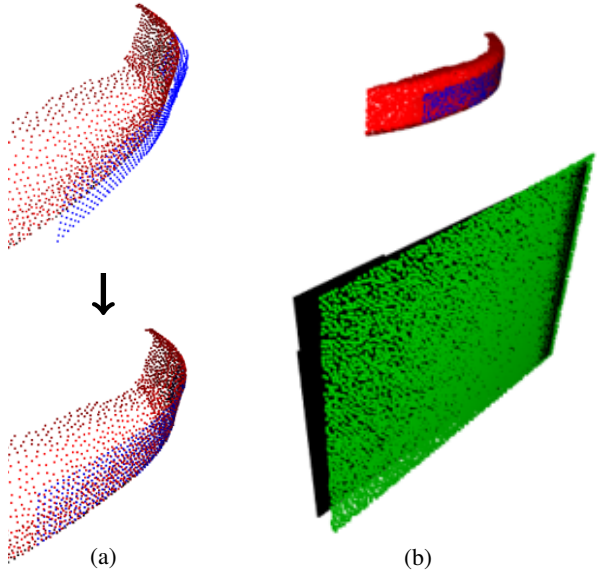

    \centering
    \inputtikz{point_cloud_alignment}
    {\phantomsubcaption\label{fig:coil_point_cloud_alignment}
     \phantomsubcaption\label{fig:misalign_point_cloud_alignment}}
    \caption{The derivation of as-tested geometry, demonstrating alignment of the 3D scan (red) and DIC derived point cloud of the surface of the induction coil (blue) (a) and the misalignment of as-designed and as-tested monoblock position used to update the simulation geometry (b).}
    \label{fig:point_cloud_alignment}
\end{figure}

We compute this mapping using point cloud representations of the as-designed and as-tested geometries.
The as manufactured induction coil differs significantly from the idealised design geometry.
We therefore used a 3D laser scanner (Nikon ModelMaker H120) to scan the induction coil, generating a mesh based representation of the outer surface.
The raw scan was repaired and smoothed, then converted to a point cloud.
We also used the same scan to construct the induction coil geometry in the simulation model.
Point cloud manipulation and mapping was implemented in Python, using the Open3D \citep{Zhou2018} and SciPy \citep{2020SciPy-NMeth} libraries.
The methods start with two point clouds, one for the as-designed geometry (using the scan-derived coil geometry) and the other for the as-tested geometry.
We first partition the as-tested point cloud using the DBSCAN clustering algorithm \citep{ester1996density} to identify two groups of nodes describing the coil and monoblock.
We coarsely align the coil point clouds using cloud centroids, which provides an adequate initial guess for Open3D's point-to-plane iterative closest point (ICP) registration algorithm, the results of which are shown in fig \ref{fig:coil_point_cloud_alignment}.
The DIC post-processing (\S\ref{subsubsec:DICPostProcessing}) is configured to capture out of plane curvature away from the front surface of the coil, which is critical in achieving robust and repeatable mapping of the point clouds using ICP registration.
The ICP registration result is described by a 4x4 transformation matrix, which we then use to transform the as-tested point cloud of the monoblock front face.
This reveals the discrepancy between the geometries in the relative position and orientation of the monoblock with respect to the coil, an example of which is shown in fig \ref{fig:misalign_point_cloud_alignment}.
We then utilise the geometric simplicity of the front face of the monoblock to compute the 4x4 transformation matrix describing this misalignment by aligning the point cloud edges and collocating the centroids.
This process is repeated for each image captured by the DIC diagnostic to provide the evolution of the geometry throughout each experiment.

\subsection{Experiment uncertainty quantification} \label{subsec:ExperimentUQ}

We implemented robust quantification of uncertainty in our experimental measurements because it drives the input characterisation within the probabilistic simulation (\S\ref{subsec:InputSpace}) and provides a means by which to isolate model discrepancy.
The aleatoric experimental uncertainty is readily computed with statistical analysis of the raw data.
Sampling at $1~\textrm{Hz}$ over a steady state period of $400~\textrm{s}$ provides sufficient data points to compute steady state statistics, assuming normal distributions for all data.
We combine the steady state data from all three repeats of the experiment and compute the statistics over the combined dataset.
There are multiple contributions to epistemic data streams.
The baseline uncertainty on all sensors, is the manufacturer reported accuracy.
All thermocouples used are k-type thermocouples, with an accuracy of the greater of $\pm 0.75\%$ or $\pm 2.2~^{\circ}\textrm{C}$.
For the thermocouples installed on the sample, we apply additional epistemic uncertainty due to the interference correction method described in \S\ref{subsubsec:Thermocouples}.
Measurement of flow conditions includes sensor accuracy derived uncertainty, which is compounded when computing heat transfer behaviour for the simulation, described in more detail in \S\ref{subsec:InputSpace}-\ref{subsec:Surrogate}.
The current and voltage measurements accumulate uncertainty through the multiple devices involved in capturing the data.
The Rogowski coil accuracy is sensitive to its position relative to the induction coil supply feedthrough, but is penalised by proximity to the return feedthrough.
The resultant compromise results in an accuracy of $\pm 6\%$ for current measurement, which is compounded by measurement uncertainty of $\pm 1.3\%$ at the PicoScope.
The voltage probe reports accuracy of $\pm 2\%$ compounded by measurement uncertainty of $\pm 2.6\%$ at the PicoScope.
Epistemic uncertainty in our DIC measurements is reduced by the DIC-levelling using the reference images captured at the start of each experiment (see \S \ref{subsubsec:DICPostProcessing}).
This renders the uncertainty in the interior of the field small relative to the aleatoric noise floor qualified below.
We note that the DIC data is less reliable at field edges, a feature discussed in \S\ref{subsec:AssessmentOfCase}.
We characterise the aleatoric uncertainty by correlating sets of 100 static images captured before and after each experiment, following the recommendations of Jones et. al \citep{jones2018}.
The translational uncertainty of our point cloud fitting has a lower bound determined by the DIC subset spacing.
This is also applied to the rotational uncertainty, but scaled by the side to side span of the point cloud to obtain a rotational uncertainty.
Geometric model inputs are therefore assumed to be epistemically accurate to $\pm 0.025~\textrm{mm}$ and $\pm 0.07^{\circ}$.

\section{Simulations} \label{sec:Simulations}
\begin{figure}
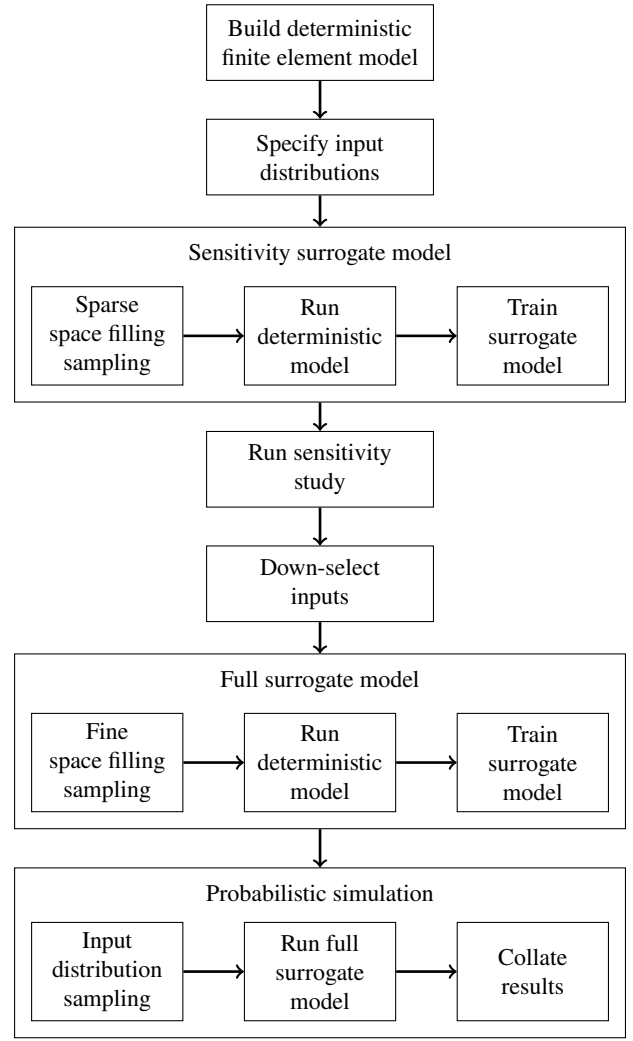

    \centering
    \inputtikz{simulation_workflow}
    \caption{The workflow for construction of a probabilistic simulation.}
    \label{fig:simulation_workflow}
\end{figure}

Our simulation work aims to model the test case as realised in the experiments described in \S\ref{sec:Experiments}.
We then demonstrate validation of this model in \S\ref{sec:Validation}.
The qualification methodology we are developing incorporates uncertainty to facilitate risk-based decision-making.
We therefore require that our simulations are probabilistic, accounting for aleatoric and epistemic uncertainties at input and predicting distributions describing system response.
We construct our probabilistic implementation as an uncertainty quantification layer above the deterministic simulation we wish to validate.
In this section we describe our deterministic finite element model (\S\ref{subsec:FEA}) and our methods to use this model to make probabilistic predictions (\S\ref{subsec:InputSpace}-\ref{subsec:ProbabilisticSimulation}).
Figure \ref{fig:simulation_workflow} shows a flowchart outlining the constituent components and process for constructing the probabilistic simulation.

\subsection{Finite element model} \label{subsec:FEA}

\begin{figure}
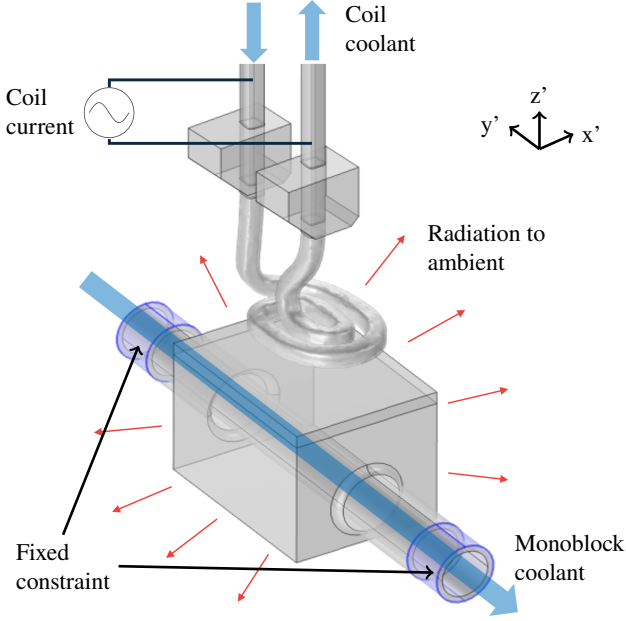

    \centering
    \inputtikz{finite_element_model}
    \caption{The finite element model, with boundary conditions annotated.}
    \label{fig:finite_element_model}
\end{figure}

The basis of all of our simulation work is a finite element model (FEM) simulating the electromagnetic-thermal-mechanical response of the system to experiment-derived boundary conditions.
This model was built and the system response solved for using the COMSOL Multiphysics\textsuperscript{\textregistered} v.6.3 software \citep{comsol}.
An overview of the model and boundary conditions is shown in Fig.~\ref{fig:finite_element_model}.

\subsubsection{Model geometry}
The model geometry combines as-designed geometry of the monoblock and machined components of the induction coil, with a thickened mesh-based geometry derived from the 3D scan (\S\ref{subsubsec:AsTestedGeometry}) of the induction coil extrusion.
This was driven by inspection of the as-tested geometry, confirming that the accuracy of machined parts was sufficient for direct use in the model, but that the manufacturing process was unable to replicate the intended coil geometry.
The geometry construction is parameterised such that the relative position and (Euler angle described) orientation of the monoblock with respect to the coil can be varied between model runs.
The model implemented for the electromagnetic simulation additionally includes a large, cuboidal vacuum domain at which to apply far-field boundary conditions.
The coil geometry is artificially extended to this domain boundary to give an open current loop with source and sink at the problem boundary.

\subsubsection{Boundary conditions}
The electromagnetic component of the simulation solves for the magnetic and electric fields (via vector and scalar potentials respectively) within all geometry, doing so in the frequency or time-harmonic domain to efficiently model the high frequency excitation of the induction coil.
The primary source term in this solve is the current density field through the induction coil, which is driven by specifying an integral current value over the coil cross-section at the problem domain boundary.
The far-field boundary condition sets the tangential components of magnetic vector potential to zero at the problem domain boundary.
The inclusion of the electric potential in the solve allows us to compute the voltage difference between two nodes in the coil, positioned to replicate the voltage probe installation in the experiment.
The time-varying nature of the magnetic field induces currents in the monoblock, and we compute the resultant volumetric electromagnetic heating in the monoblock as well as the direct electromagnetic heating in the coil due to the excitation current.

The thermal component of the simulation solves for the steady state temperature field in the monoblock and coil.
The spatially varying, time-averaged electromagnetic heating field output from the electromagnetic solve is applied as a volumetric heat source.
The two coolant flows provide two heat sinks to the system, which are modelled as surface convection heat fluxes for which we specify a coolant temperature and heat transfer coefficient.
The temperature of both the coil and the monoblock is sufficiently high that the heat flux due to radiation to background is non-negligible.
This is modelled as a surface heat flux assuming radiation from each surface to a black body background.
In addition to the full temperature field output, we also report local values of temperature at locations matching the thermocouple installation in the experiment.

The mechanical component of the simulation solves for the displacement field within the monoblock, modelling the thermal expansion of the component due to the temperature field output from the thermal solve.
Additional loading is applied to represent the monoblock coolant pressure.
The system is constrained by requiring zero displacement and rotation over the outer surfaces of the pipe at which the monoblock interfaces with the HIVE coolant system in the experiment (it is noted that the experimental system is not completely stiff).
We output strain and displacement values over the front face of the monoblock, matching the field of view of the DIC diagnostic in the experiment.

The majority of boundary condition inputs can be extracted directly from the time-averaged steady state experimental data (or a sample from an input uncertainty distribution).
However, the heat transfer coefficients used are calculated from multiple experiment outputs.
We compute heat transfer coefficients by considering the mean flow in each coolant system.
Using the monoblock coolant condition as an example, we can compute mean velocity, $\bar{U}$, from the experimental coolant flow rate, $Q$, as
\begin{equation}
    \bar{U} = \frac{4Q}{\pi D^2},
\end{equation}
where $D$ is the internal pipe diameter.
We can then compute the Reynolds number, $\textrm{Re}$, of the flow as
\begin{equation}
    \textrm{Re} = \frac{\rho \bar{U} D}{\mu},
\end{equation}
where $\rho$ and $\mu$ are the fluid density and viscosity respectively.
The typical flow rates of the monoblock and coil coolant flows give Reynolds numbers of $\sim 4 \times 10^5$ and $\sim 3 \times 10^4$ respectively, and we can therefore assume both flows are turbulent.
We use the Dittus-Boelter equation \citep{dittus1985heat} to estimate a heat transfer coefficient for turbulent pipe flow.
Dittus-Boelter computes the Nusselt number, $\textrm{Nu}$, from the Reynolds number and Prandtl number, $\textrm{Pr}$, as
\begin{equation}
    \textrm{Nu} = 0.023 Re^{0.8} Pr^{0.4},
\end{equation}
where
\begin{equation}
    \textrm{Pr} = \frac{c_p \mu}{k},
\end{equation}
with the fluid specific heat capacity, $c_p$, and thermal conductivity, $k$.
We then use the definition of the Nusselt number
\begin{equation}
    \textrm{Nu} = \frac{h D}{k}
\end{equation}
to compute a heat transfer coefficient, $h$.
Dittus-Boelter is derived for circular cross-section pipes assuming small temperature difference between bulk fluid and pipe surface.
This is a good approximation to the monoblock coolant boundary condition, but less good for the coil, for which the coolant channel has a rounded square cross-section and the fluid-solid temperature difference is greater.
We account for this by the addition of significant epistemic uncertainty to the coil heat transfer input in the probabilistic simulation (see \S\ref{subsec:ProbabilisticSimulation}).

\subsubsection{Material properties}
We implemented temperature dependent material properties for the Oxygen-Free High Conductivity (OFHC) Copper coil and 316L stainless steel monoblock, with source data from UKAEA's materials database.
Electromagnetic properties for both materials were modelled as isotropic, with temperature dependent electrical conductivity and relative magnetic permeability $\mu_r = 1$ (no material magnetisation).
The thermal properties were also modelled as isotropic, with temperature dependent thermal conductivity, specific heat capacity, and density.
The surface emissivities of the materials are assumed to be constant over the temperature ranges seen in the experiments.
The mechanical properties of the monoblock were modelled as linear elastic and isotropic with a temperature dependent elastic modulus and coefficient of thermal expansion, and constant Poisson's ratio.

We parameterise the material properties to allow variation within the probabilistic simulation.
This is implemented by applying a scalar to each property of interest, including to the expressions describing the temperature dependence of certain properties.
Usage of this parameterisation is described in \S\ref{subsec:InputSpace}.

\subsubsection{Numerical representation and solve structure}

Our numerical representation of the problem posed above uses ten-node tetrahedral meshes, with second order (quadratic) elements.
The electromagnetic solve uses an A-V formulation, simultaneously solving for magnetic and electric field behaviour.
Since the excitation of the induction heating coil is time harmonic, the electromagnetic problem can be solved in the frequency domain, with frequency set at the coil excitation frequency from the experiment.
Electromagnetic heating is averaged over a full cycle of the frequency domain solve to provide a steady state heat source input to the thermal solve.
The linear solve of the electromagnetic system is performed using the GMRES iterative solver.
For the thermal problem we solve directly for the temperature field in a steady state solve.
The linear solve within the thermal problem is performed using the GMRES iterative solver.
We perform a steady state solve of the solid displacement vector to compute the mechanical response of the monoblock, assuming small displacement behaviour.
We use the PARDISO direct solver for the linear mechanical solve.
The displacement field is used to compute the strains on the front face of the monoblock, assuming small strains.

We construct the overall solve to suit the sensitivity of the various multi-physics couplings and to facilitate comparison with the experiment.
The electromagnetic and thermal solves were found to be two-directionally sensitive, and were therefore solved using a tight co-simulation approach.
This involves an additional iterative loop at a higher level than the individual solves, in which updated temperature and electromagnetic heating fields are passed between the two solvers at each iteration.
The displacements output from the mechanical solve were small, sufficiently so that the influence on the electromagnetic solve is negligible.
The mechanical solve was therefore run separately as the final solve step, taking the converged temperature field from the electromagnetic-thermal co-simulation as input.
All high-level solve steps (electromagnetic-thermal, and mechanical) are non-linear.
The electromagnetic-thermal solve must resolve the effects of the co-simulation as well as the temperature dependence of the material properties.
We use a Newton-Raphson method to perform the non-linear solve accounting for both of the above contributions to non-linearity.
The mechanical solve must account for temperature dependence of the material properties, and also uses a Newton-Raphson method.

\begin{figure}
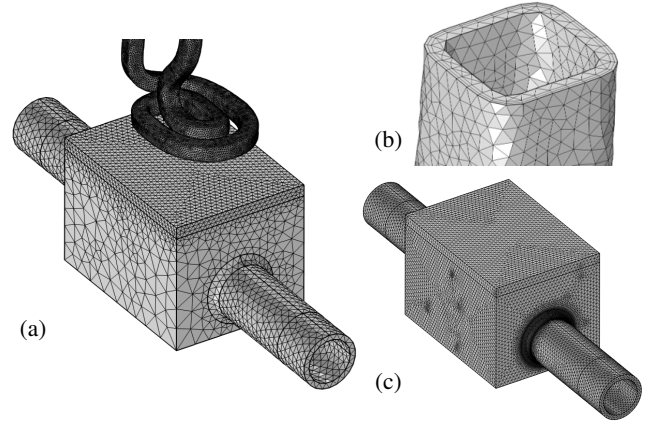

    \centering
    \inputtikz{mesh}
    {\phantomsubcaption\label{fig:mesh_EM}
     \phantomsubcaption\label{fig:mesh_coil}
     \phantomsubcaption\label{fig:mesh_ST}}
    \caption{Key features from the finite element model mesh. (a) shows the electromagnetic solve mesh, with the detail of the coil skin depth meshing shown in (b). The mesh for the mechanical solve is shown in (c).}
    \label{fig:mesh}
\end{figure}

To build the numerical mesh we employ physics driven meshing strategies, followed by refinement to achieve solution convergence with respect to the mesh (convergence reported in \S\ref{subsubsec:ModelVerification})
The demands on meshing differ between the different physics being solved in our problem.
We utilise the separation of solves in the structure described above to allow use of two meshes to suit the different components of the solve.
In a frequency-dependent magnetic field, as we have in our system, the spatial behaviour of currents, both direct and induced, is governed by skin effects \citep{hayt2006engineering}.
Skin effects result in exponential decay of the current density field, with the real part described by
\begin{equation}
    \mathbf{J} = \mathbf{J_s} e^{-\frac{d}{\delta}},
\end{equation}
where $\mathbf{J}$ is the current density vector at some depth into the material $d$, $\mathbf{J_s}$ is the current density vector at the surface, and $\delta$ is the problem skin depth.
Skin depth is dependent on the material supporting the current density field and the temporal behaviour of the field, with
\begin{equation}
    \delta = \frac{1}{\sqrt{\pi \mu \sigma f}},
\end{equation}
where $\mu$ is the material magnetic permeability, $\sigma$ the electrical conductivity, and $f$ is the characteristic frequency of the changing magnetic field.
For our problem, we have two values for skin depth, one for copper (the coil) and one for stainless steel (the monoblock), due to the difference in electrical conductivity.
In the finite element model, we require adequate resolution of the skin depth, which we implement as at least two elements through the skin depth thickness.
We define a region for skin depth meshing behind the upper face of the monoblock, shown in Fig.~\ref{fig:mesh_EM}, and utilise boundary layer meshing applied from the outer surface of the coil, as shown in Fig.~\ref{fig:mesh_coil}.
The mechanical solve only models the behaviour of the monoblock, and therefore does not require meshing of the coil and vacuum domains.
It also does not require, and is wasteful to solve with, the skin depth meshing implemented in the electromagnetic-thermal solve.
It instead requires sufficient mesh resolution to resolve thermal gradients, and increased resolution at stress raising geometric features, which for the monoblock are the fillets between the main block and the pipe.
The resultant mesh is shown in Fig.~\ref{fig:mesh_ST}.
We therefore separate the meshing of the electromagnetic-thermal and mechanical solves to suit the requirements described above.
Information transfer between the meshes is limited to the temperature field in the monoblock, which, as a non-conservative quantity, can be accurately mapped between meshes using an interpolation method that evaluates temperature using the second order shape function of the electromagnetic-thermal mesh elements.

The problem geometry is defined at room temperature, the temperature at which the components were manufactured and their dimensions measured.
However, the DIC reference image is taken when the sample is at coolant temperature.
Therefore, the DIC measured thermal strains are zero at the coolant temperature.
In order to replicate the relative displacements (and thereby strains) the DIC diagnostic measures, a thermal solve and a subsequent mechanical solve are also performed, but with the thermal system given no electromagnetic heat input.
The simulated DIC results are given as the difference between the full solution and the no-heat-load solution.

\begin{figure*}
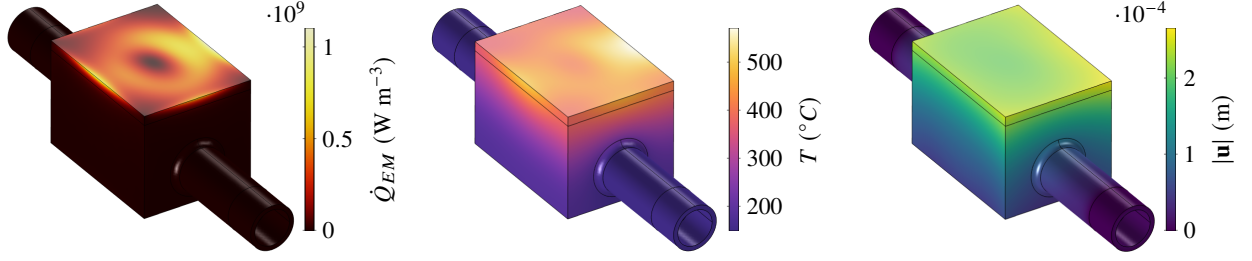

    \centering
    \inputtikz{full_field_sim_results}
    \caption{Finite element simulation results for nominal input parameter values, plotting the electromagnetic heating, the temperature field, and the solid displacement of the monoblock.}
    \label{fig:full_field_sim_results}
\end{figure*}

\begin{figure*}
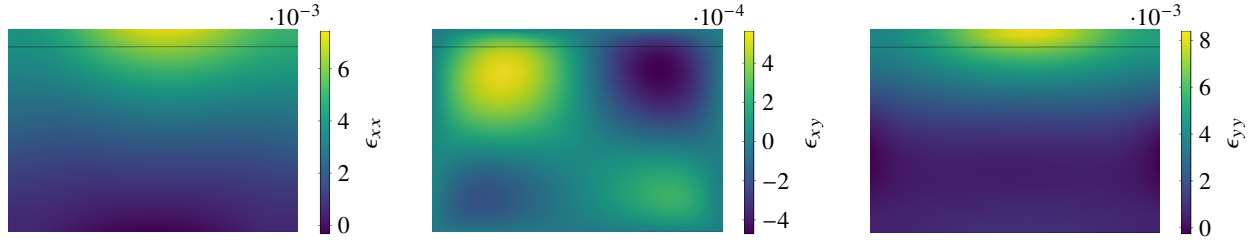

    \centering
    \inputtikz{nominal_sim_strains}
    \caption{Finite element simulation results for nominal input parameter values, plotting in-plane components of the strain tensor on the camera view face.}
    \label{fig:nominal_sim_strains}
\end{figure*}

Deterministic predictions of the electromagnetic heating, temperature, and solid displacement fields for the steady state response of the sample are shown in Fig.~\ref{fig:full_field_sim_results}.
It can be seen that despite efforts to smooth the heating by induction coil design and reduced coil proximity, there is significant spatial variation in the field.
The effects of this are evident in the temperature field, but are less visible in the displacement field where global thermal strain is dominant.
The resultant strain fields on the camera view face are shown in Fig.~\ref{fig:nominal_sim_strains}.

\subsubsection{Model verification} \label{subsubsec:ModelVerification}
A compulsory step before performing any model validation against experiment is verification of the model setup and solution \citep{oberkampf2010verification}.
We report on two key aspects of this for our finite element model: mesh convergence and solution convergence.

\begin{figure}
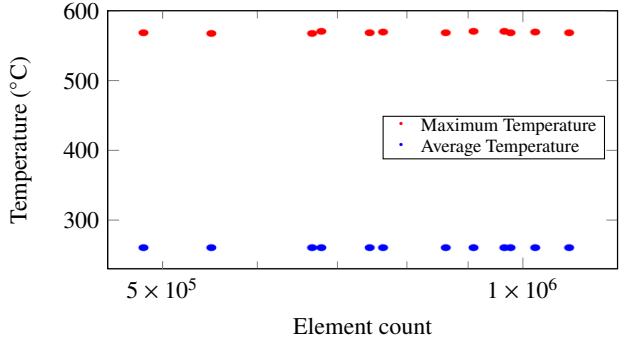
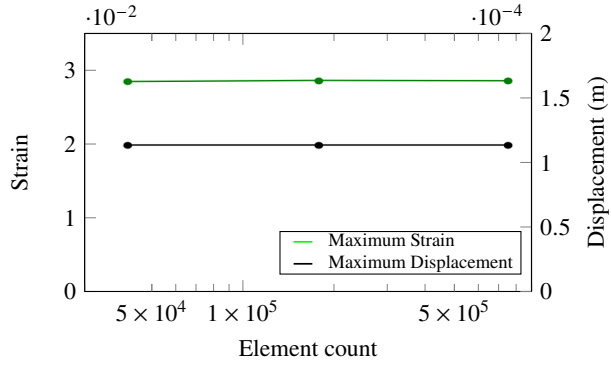

    \centering
    \begin{subfigure}{\linewidth}
        \centering
        \inputtikz{mesh_convergence_EM}
        \caption{Electromagnetic-thermal solve mesh. Plotted without connecting lines due to variation of multiple independent mesh parameters across data points.}
        \label{fig:mesh_convergence_EM}
    \end{subfigure}
    \begin{subfigure}{\linewidth}
        \vspace{0.25cm}
        \centering
        \inputtikz{mesh_convergence_struct}
        \caption{Mechanical solve mesh.}
        \label{fig:mesh_convergence_struct}
    \end{subfigure}
    \caption{Mesh convergence plots for both meshes used in the solve. The electromagnetic-thermal mesh has multiple refinement controls, all of which are varied throughout the convergence study, and hence more datapoints in the convergence study than for the mechanical mesh.}
    \label{fig:mesh_convergence}
\end{figure}

We confirm suitability of our meshing by performing mesh convergence studies, varying the resolution of meshes to confirm invariance of model predictions with mesh resolution at the resolution used to generate our results.
This is done for all aspects of the mesh refinement of each of the meshes, for example for the skin depth meshing in the electromagnetic-thermal mesh, to ensure that all mesh related behaviour has converged.
The convergence results are plotted in Fig.~\ref{fig:mesh_convergence}, and demonstrate converged behaviour significantly below the mesh resolution used, most likely due to the use of second order shape functions.
We retain high mesh resolution to maintain the accuracy of post-processing of field data outside of the finite element model.
The final meshes used consist of $1.02 \times 10^6$ elements for the electromagnetic-thermal solve, and $7.5 \times 10^5$ elements for the mechanical solve.

We must confirm that the process used to solve for component response in each of the physics domains of interest has adequately converged to a stable solution.
During the solve, the solution error is computed and compared to a prescribed tolerance.
All solves must achieve a scaled relative residual below a tolerance value of $10^{-3}$.
The scaling is used to account for how well conditioned the system is, such that relatively ill-conditioned systems see a larger number of iterations.
For the mechanical solve, which is based on a direct linear solve, the relative residual is used unscaled.
Our iterative thermal solve, which has a single degree of freedom per node and only solves on the mesh within the coil and monoblock, uses a residual scaling of $20$ and therefore an effective tolerance of $5 \times 10^{-5}$.
The electromagnetic solve is much larger, with four degrees of freedom per node and additionally solving over the full vacuum domain, and therefore uses a residual scaling of $10^4$ and an effective tolerance of $10^{-7}$.
If any of these error checks are not satisfied the simulation fails with no solution.
We can therefore confirm that all solutions satisfy our solve accuracy requirements.

In our solve structure, individual solver convergence is not the only check we must make.
We model the electromagnetic-thermal behaviour as coupled, and therefore perform a co-simulation of these two solves.
A stopping criterion is applied on the co-simulation iterations based on the relative error in the solutions to each individual solve, continuing until the relative change in solution over an iteration compared to that of the first iteration falls below a tolerance of $10^{-3}$.
Again, if this condition is not met the simulation fails with no solution, and we are therefore able to confirm that the coupled solutions generated are acceptable.

\subsection{Input parameter space and sensitivity study} \label{subsec:InputSpace}
We deliberately varied the inputs to our deterministic model (geometry, boundary conditions and material properties) to understand the sensitivity of the system to input uncertainty in our simulation.
This is used in a sensitivity study to down select inputs to those with significant influence on the system, and in the final probabilistic simulation.
The experimental data informs the variation of geometric inputs, defining the relative position and rotation of the monoblock with respect to the induction coil, and any directly measured boundary conditions (coolant temperature and pressure, and coil excitation).
The ranges are a combination of measured aleatoric uncertainty and diagnostic-accuracy-derived epistemic error.
Unmeasured inputs, such as the remaining boundary conditions (coolant heat transfer, and solid surface emissivity) and material properties must also be varied, but the range of input values is a conservatively assigned epistemic uncertainty.
The inputs assessed are recorded in Table \ref{tab:input_uncertainties}, along with their assigned uncertainties and down-selection status.
The aleatoric uncertainty from our experimental data is fitted to normal distributions.

\begin{table*}
    \centering
    \begin{tabular}{|l|l|l|l|l|l|}
        \hline
        \textbf{Input} & \textbf{Unit} & \textbf{Mean value} & \textbf{Aleatoric} & \textbf{Epistemic} & \textbf{Probabilistic input} \\
        \hline
        Coil current & A & 1550.0 & 0.46 & \textbf{$\pm$94.2} & Yes\\
        Coil frequency & kHz & 123.5 & 0.038 & \textbf{$\pm$0.5} & Yes \\
        Misalignment angle, $\alpha$ & deg & 0.70 & \textbf{0.47} & $\pm$0.07 & Yes \\
        Misalignment angle, $\beta$ & deg & -0.32 & \textbf{0.35} & $\pm$0.07 & Yes \\
        Misalignment angle, $\gamma$ & deg & -3.86 & \textbf{0.26} & $\pm$0.07 & No \\
        Misalignment, dx' & mm & -1.42 & \textbf{0.26} & $\pm$0.025 & Yes \\
        Misalignment, dy' & mm & 0.50 & \textbf{0.30} & $\pm$0.025 & Yes \\
        Misalignment, dz' & mm & -0.23 & \textbf{0.26} & $\pm$0.025 & Yes \\
        Monoblock coolant temperature & $^\circ$C & 149.0 & 0.51 & \textbf{$\pm$2.2} & Yes \\
        Monoblock coolant HTC & $\textrm{W m}^{-2}\textrm{K}^{-1}$ & 43600 & 0 & \textbf{$\pm$4300} & Yes \\
        Coil coolant temperature & $^\circ$C & 45.2 & 0.38 & \textbf{$\pm$2.2} & No \\
        Coil coolant HTC & $\textrm{W m}^{-2}\textrm{K}^{-1}$ & 27800 & 0 & \textbf{$\pm$5500} & No \\
        Monoblock emissivity & & 0.3 & 0 & \textbf{$\pm$0.05} & No \\
        Coil emissivity & & 0.5 & 0 & \textbf{$\pm$0.05} & No \\
        Speckle emissivity & & 0.9 & 0 & \textbf{$\pm$0.05} & No \\
        Thermal expansion, steel & \% & 100 & 0 & \textbf{$\pm$5} & Yes \\
        Electrical conductivity, steel & \% & 100 & 0 & \textbf{$\pm$5} & Yes \\
        Thermal conductivity, steel & \% & 100 & 0 & \textbf{$\pm$5} & Yes \\
        Young's modulus, steel & \% & 100 & 0 & \textbf{$\pm$5} & Yes \\
        Poisson's ratio, steel & \% & 100 & 0 & \textbf{$\pm$5} & Yes \\
        Specific heat capacity, steel & \% & 100 & 0 & \textbf{$\pm$5} & Yes \\
        Density, steel & \% & 100 & 0 & \textbf{$\pm$5} & No \\
        Electrical conductivity, copper & \% & 100 & 0 & \textbf{$\pm$5} & No \\
        Specific heat capacity, copper & \% & 100 & 0 & \textbf{$\pm$5} & No \\
        Thermal conductivity, copper & \% & 100 & 0 & \textbf{$\pm$5} & Yes \\
        Density, copper & \% & 100 & 0 & \textbf{$\pm$5} & Yes \\
        Coolant pressure & Pa & 610500 & 460 & \textbf{$\pm$11000} & No \\
        \hline
    \end{tabular}
    \caption{The contributions to overall input uncertainty from aleatoric and epistemic uncertainty. The aleatoric column records the standard deviation of a normal distribution describing the uncertainty. We utilise the fact that one type of uncertainty dominates each input (highlighted in bold) to only implement that single uncertainty type in the simulation.}
    \label{tab:input_uncertainties}
\end{table*}

The number of inputs implemented within a probabilistic simulation dictates the dimensionality of the sampling problem.
At higher dimensionality the sampling problem becomes extremely large, with the number of samples typically increasing exponentially with number of dimensions in a phenomenon known as the curse of dimensionality \citep{donoho2000high}.
Similarly, the amount of training data required to build an accurate surrogate model (discussed in \S\ref{subsec:Surrogate}) increases with number of inputs.
To make both of the above feasible, we must only model variation of inputs that have tangible impact on the results.
We therefore perform a sensitivity study on our input space to down-select to the input set that we take forward into the remainder of the simulation process.

We run a sensitivity analysis using the SmartUQ\textsuperscript{\textregistered} v.10.1.1 software package \citep{smartuq}.
The method used interrogates the variance of model outputs with respect to variance in the inputs and attributes fractions of the output variance to each input \citep{sobol2001global}.
The analysis computes ``main effect" and ``total effect" indices for each input, which describe the contribution of an input to output variance when the input is considered in isolation and in combination with all other inputs respectively.
Computation of relevant output and input variance requires sampling over the input space of all parameters to be assessed, and execution of the forward predictive model for each sample, the very problem we are trying to mitigate by performing the sensitivity study.
However, we are here sampling from input ranges rather than distributions, and can still obtain useful results for input down-selection from coarse sampling.
To further improve the efficiency of the sensitivity study, we use a surrogate model, the ``sensitivity surrogate" (see \S\ref{subsubsec:SurrogateMethodsTraining} for details), as our forward predictor.
Table \ref{tab:input_uncertainties} reports the uncertainties on 27 inputs to the model, which the sensitivity study must assess.
We use Sobol sequence \citep{sobol1967distribution} based sampling, which allows for quasi-random sampling with improved uniformity to adequately span the input space with relatively fewer samples than purely random sampling.
The Sobol sequence generates 2700 samples.
We run the sensitivity analysis probabilistically, performing 10 repeats of the Sobol sequence sampling and therefore a total of 27000 executions of the sensitivity surrogate.
This allows us to compute statistics on effect indices over the repeats and thereby confirm that our sparse sampling approach is not introducing unacceptable uncertainty.
The sensitivity surrogate predicts the thermocouple temperatures, coil voltage, and values of each of the modal coefficients used to reconstruct the strain fields (see \S\ref{subsec:Surrogate} for further details).

\begin{figure*}
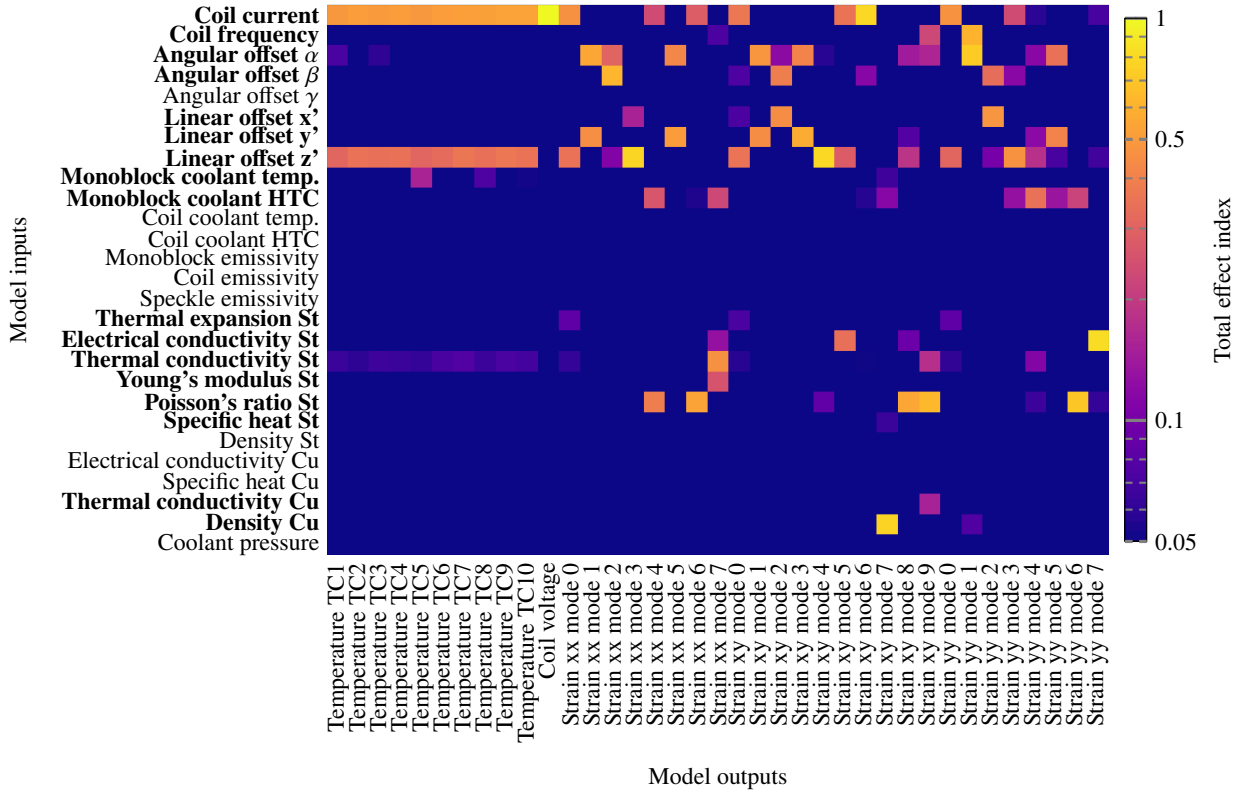

    \centering
    \inputtikz{sensitivity}
    \caption{The results from the sensitivity study used to down-select inputs for use in the probabilistic simulation, plotting total effect indices. A threshold of 5\% was applied to the maximum total effect index of each input for inclusion in the probabilistic simulation. This threshold is shown by the lower limit of the colour scale in the plot. Inputs taken forward for the probabilistic simulation are highlighted in bold.}
    \label{fig:sensitivity}
\end{figure*}

We assess the sensitivity to inputs in our problem using the total effect indices.
The results from the sensitivity study are shown in Fig.~\ref{fig:sensitivity}.
The sensitivity of our test case to the relative position and orientation of the monoblock and induction coil is evident, motivating the post-processing described in \S\ref{subsubsec:AsTestedGeometry}, as is the importance of the coil current.
We apply a 5\% threshold on maximum total effect index to select the inputs to be used in the probabilistic simulation.
This negates a number of material property variations, unmeasured boundary conditions such as surface emissivity, and some of the measured boundary conditions such as monoblock coolant pressure and coil coolant conditions.

\subsection{Surrogate modelling} \label{subsec:Surrogate}
Performing probabilistic simulation requires sampling from input distributions and forward predictive model execution at each sample point.
Adequate resolution of the input domain therefore requires a large number of executions of the model ($\mathcal{O}(10^5)$).
Completion of a single simulation using the FEA model described in \S \ref{subsec:FEA} requires $\mathcal{O}(2~\textrm{hr})$ compute time on a 32-core cluster node, using Intel\textsuperscript{\textregistered} Xeon\textsuperscript{\textregistered} Gold 5218 CPUs running at $2.30~\textrm{GHz}$ and requiring $130~\textrm{GB}$ of physical memory.
Running 1000s of executions of this model is therefore intractable.
We therefore used surrogate models to allow us to perform the required sampling 
for our sensitivity study and probabilistic simulation.
There are two main strategies for creating surrogate models: 1) simplification of the model physics; or 2) statistical fitting of a suitable function between model inputs and outputs.
We used the second approach with a Gaussian process as the surrogate model.
We require two separate surrogates within our simulation workflow: a ``sensitivity surrogate" to allow efficient down-selection of inputs via a sensitivity study, and a ``full surrogate" to predict all finite element model outputs from variation of the down-selected input set.
Both surrogates predict full field strain response of the monoblock in addition to the point sensor predictions.
However, the sensitivity surrogate uses all inputs in Table \ref{tab:input_uncertainties}, with fewer training samples and is therefore less accurate.

\subsubsection{Surrogate methods and training} \label{subsubsec:SurrogateMethodsTraining}
For each surrogate model built, we required two sets of data from the finite element model: training data and independent validation data.
Training of the sensitivity surrogate required 50 evaluations of the finite element model, with an additional 5 evaluations for validation.
Our accuracy requirements for the full surrogate are more stringent, and we used 209 evaluations for training and 24 evaluations for validation.
We used Latin Hypercube sampling \citep{mckay2000comparison} over the relevant ranges of the input parameters to efficiently generate independent Design of Experiments (DOEs) for each of the datasets described above all of which span the full input space.
The input ranges used are the same as those used in the sensitivity study, described in \S\ref{subsec:InputSpace}.
Each of the finite element model evaluations outputs the temperatures at each of the thermocouple locations and the voltage across the induction coil voltage probe, as well as mesh based 2D datasets of in-plane components of strain on the front face of the monoblock.

To replicate the output of our finite element model, we use a Gaussian Process surrogate model \citep{williams2006gaussian} defined, trained, and run using the SmartUQ\textsuperscript{\textregistered} software package and Python API.
Choice of kernel is an important initial step in Gaussian Process training.
Our nominal simulation results, shown in Fig.~\ref{fig:full_field_sim_results}, show fields that are smooth but with significant peaks towards the edges of the top face, particularly in the electromagnetic solution.
It follows that the responses that the surrogate must model will have similar characteristics in response to variation at input.
For example, global response of a thermocouple close to the top face of the monoblock will be generally smooth, but see steeper gradients for input conditions that move it close to the induction heating coil.
We therefore use a Mat\`{e}rn kernel, which gives a smooth response, but with the ability to resolve moderate gradients.
The Gaussian process surrogate predicts single value outputs, and is poorly suited to prediction of full field data.
We therefore combine this surrogate with a field decomposition approach to allow prediction of strain fields.

\begin{figure*}
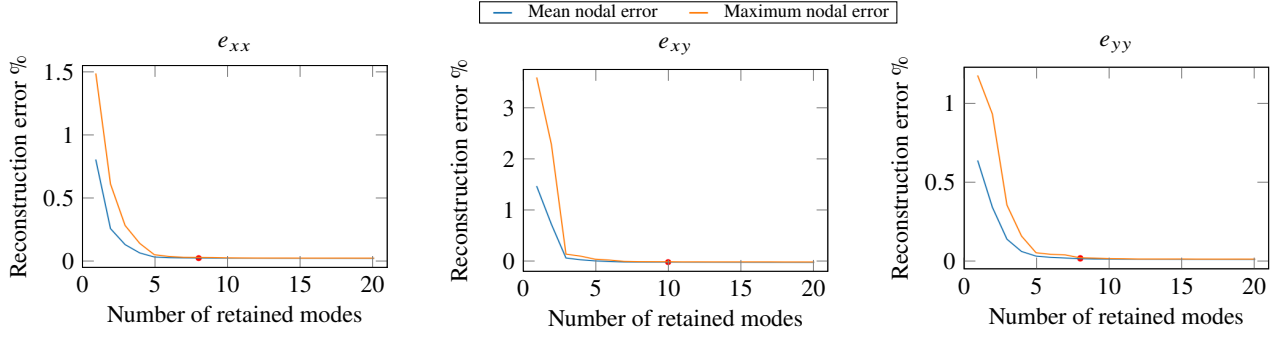

    \centering
    \inputtikz{modal_decomposition_error}
    \caption{The variation in the error introduced by model reconstruction of strain fields against the number of modes used. The red dot indicates the number of modes used in the surrogate models.}
    \label{fig:modal_decomposition_error}
\end{figure*}

The variability of field data generated from spanning our input parameter space is relatively linear, and we can therefore use a modal decomposition approach, singular value decomposition (SVD), to characterise field behaviour.
We perform the SVD factorisation and subsequent matrix operations using the linear algebra functions in the NumPy Python library \citep{harris2020array}.
To implement SVD we require our field data to be on a common mesh, which is not the case in the raw exports from the finite element model due to remeshing for each DOE design.
We interpolate all field data onto a regular grid, with a monoblock face oriented coordinate system centred at the face centroid, using a piecewise cubic and continuously differentiable interpolator.
To perform SVD, the interpolated fields are combined into a matrix $\mathbf{M}$ with shape $n_{nodes} \times n_{fields}$, with each field collapsed into a single column.
The decomposition describes $\mathbf{M}$ as
\begin{equation}
    \mathbf{M} = \mathbf{U \Sigma V^T},
\end{equation}
where $\Sigma$ is a diagonal matrix with the singular values on the diagonal, and $\mathbf{U}$ and $\mathbf{V}$ are matrices of the left singular vectors and right singular vectors respectively.
The columns of $\mathbf{U}$ have the same length as the collapsed training fields, and are the mode shapes that describe the behaviour of the training field set.
SVD can be performed such that the singular values are ordered in descending order, in which case the mode shapes are ordered from most to least important in their description of overall field behaviour.
Any training field, $\mathbf{A}$, can be exactly reconstructed from the product of the mode shapes, $\mathbf{U}$, and corresponding mode coefficients, $\mathbf{\sigma}$, as
\begin{equation}
    \mathbf{A} = \mathbf{U \sigma},
\end{equation}
where $\mathbf{\sigma}$ is a diagonal matrix with the coefficients on the diagonal.
Mode coefficients can therefore be computed for each training field as
\begin{equation} \label{eq:mode_coefficients}
    \mathbf{\sigma} = \mathbf{A^T U}.
\end{equation}
We utilise the ordering of mode shapes in the SVD to compute an approximate reconstruction of each training field, $\tilde{\mathbf{A}}$, using only the first $n$ mode shapes, $\tilde{\mathbf{U}}$.
This allows a significant reduction of the dimensionality of the reconstruction, making the reconstruction approach suitable for a fast-executing surrogate model.
The value of $n$ is a compromise between reconstruction accuracy and dimensionality, and is selected at the point of diminishing returns in accuracy improvement.
This is demonstrated in Fig.~\ref{fig:modal_decomposition_error}, with selection of 8, 10, and 8 modes for $e_xx$, $e_xy$, and $e_yy$ strain fields respectively.

The overall methodology facilitates generation of predicted fields from a set of invariant mode shapes combined with prediction specific mode coefficients.
The mode coefficients are therefore suitable for prediction by our Gaussian Process surrogate model.
We train our Gaussian Process surrogates using the first $n$ mode coefficients computed from (\ref{eq:mode_coefficients}), $\tilde{\mathbf{\sigma}}$, for each training field.
Our hybrid surrogate models can therefore predict field data by first predicting the mode coefficients for each field type, and then reconstructing the full field by combining with the precomputed mode shapes.

\subsubsection{Surrogate model validation}
\begin{figure*}
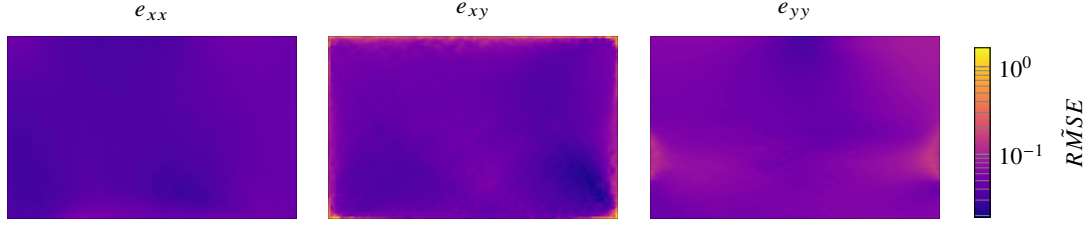

    \centering
    \inputtikz{surrogate_strain_srmse}
    \caption{Standardised root mean square error of strain field surrogate predictions compared with finite element validation data.}
    \label{fig:surrogate_strain_srmse}
\end{figure*}

Having trained the surrogate model, we then validate that it adequately replicates the finite element model using the validation data set kept separate from the training data set.
For this set of input definitions we compare surrogate model predictions, $\hat{y}_i$, with finite element predictions, $y_i$, computing root mean squared error (RMSE) between the two, normalised by the standard deviation of finite element results over the sample space, $\sigma$, as
\begin{equation}
    \textrm{RMSE}^\prime = \frac{\sqrt{\frac{1}{n}\sum_{i=1}^{n}\left(\hat{y}_i - y_i\right)^2}}{\sigma}.
\end{equation}
This provides a metric describing the importance of the error when compared to the operational range of the surrogate, with $\textrm{RMSE}^\prime << 1$ indicating acceptable surrogate prediction.
The maximum $\textrm{RMSE}^\prime$ value over all single values output by the simulation was $0.054$, reported for one of the thermocouples.
For validation of strain field data, our surrogate predictions are the reconstructed field results, thereby incorporating Gaussian Process error in mode coefficient prediction and error associated with the decomposition and modal reduction.
We compute $\textrm{RMSE}^\prime$ pointwise over the in-plane strain fields, plots of which are shown in Fig.~\ref{fig:surrogate_strain_srmse}.
The maximum $\textrm{RMSE}^\prime$ values for the $e_{xy}$ and $e_{yy}$ surrogate predictions are seen to be larger than acceptable, at $1.08$ and $0.12$ respectively.
However, when cross-referenced against the absolute strain values in Fig.~\ref{fig:nominal_sim_strains} we note that these peaks correspond with regions with very low absolute strain.
Additionally, the mean $\textrm{RMSE}^\prime$ values over the fields are acceptable at $0.025$ and $0.011$ ($e_{xy}$ and $e_{yy}$ respectively).
We therefore assume that the maximum $\textrm{RMSE}^\prime$ values are an artefact of the normalisation by standard deviation.

The surrogate validation errors reported above were considered acceptable for this study, and informed the decision to not generate additional training data.
The errors contribute an additional source of epistemic uncertainty to the probabilistic predictions, added as the final step in output processing.

\subsection{Probabilistic simulation} \label{subsec:ProbabilisticSimulation}

The probabilistic simulation can be summarised as sampling from input distributions, propagation of each sample through a forward predictive model, and compilation of output probability distributions from the model results.

We must sample from aleatoric and epistemic input uncertainty, as defined in Table \ref{tab:input_uncertainties}.
Sampling from aleatoric distributions involves simultaneous random generation biased by the distribution shape, a normal distribution for all of our experiment-derived aleatoric uncertainty.
This sort of sampling is commonplace in any Monte Carlo based simulation methodology.
We use Monte Carlo sampling to draw 400 aleatoric samples; a sample count that generates smooth output cumulative distribution functions (CDFs).
Sampling an input with epistemic uncertainty requires a different approach, accounting for the lack of knowledge of the structure of the input space. 
Our knowledge of inputs with epistemic uncertainty is limited to an upper and lower bound of the input space.
What we require at output of the simulation is epistemic uncertainty with the widest possible range, given all input epistemic uncertainty.
The input combination that achieves this is not guaranteed to have input values at an extreme of their epistemic range.
We therefore perform a grid search within all epistemic input space to identify the combination of input values that gives the minimum and maximum epistemic bound of each output.
We do so by uniform, space-filling sampling within each epistemic uncertainty range, and identification of the extreme results at output after all sampling and model execution has been completed.
We draw 250 epistemic samples; a sample count chosen using a manual convergence study terminated at the point of diminishing change in the ranges of system output values. 
As our input space includes both aleatoric and epistemic uncertainty, we must implement both sampling methodologies to generate our input sets for simulation.
The fundamental differences in our representation of these two uncertainty types requires that sampling is independent, and as such we use nested sampling where all aleatoric sampling is repeated for each epistemic sample.
As with any nested loop, this requirement dramatically increases the amount of computation required to run the probabilistic simulation.
Our choice of 400 aleatoric samples and 250 epistemic samples results in a total sample count of $10^5$.

The approach outlined above assumes that no input has both aleatoric and epistemic uncertainty, which is not the case in our experiments, as recorded in Table \ref{tab:input_uncertainties}.
It is possible to apply both uncertainty types to a single input, an example method being application of Dempster-Shafer theory \citep{dempster1968generalization,shafer2020mathematical} in which multiple epistemic intervals are assigned a probability derived from the aleatoric distribution of that input.
However, the nature of our input space allows us to avoid this additional complexity by assigning an input with one or other of aleatoric or epistemic uncertainty.
We are able to do so because each of our inputs is dominated by one of the uncertainty types.
We compare the width of two standard deviations from the aleatoric distribution with the half-range of the epistemic uncertainty, and identify the dominant uncertainty type highlighted for each input in Table \ref{tab:input_uncertainties}.

\begin{figure}
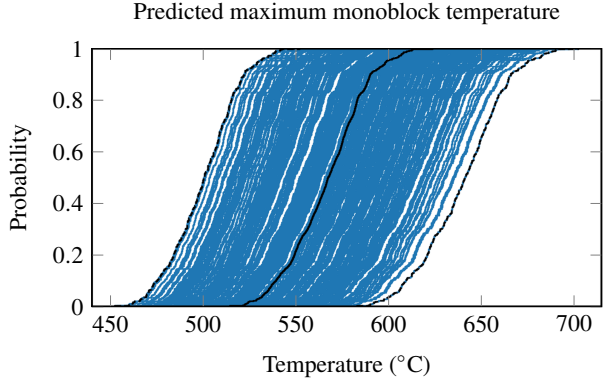

    \centering
    \inputtikz{probabilistic_max_temperature}
    \caption{The probabilistic prediction of maximum temperature within the monoblock. Each CDF represents a single epistemic sample, itself including a full set of aleatoric samples.}
    \label{fig:probabilistic_max_temperature}
\end{figure}

Having generated our samples, we execute the full surrogate model (\S\ref{subsec:Surrogate}) to generate output data.
The structure of the results mirrors the nested sampling approach, with each epistemic sample represented by a CDF made up of results from sampling of the aleatoric uncertainty.
The result for a given output is therefore an ensemble of CDFs, the bounds of which form the probability box describing the probabilistic prediction.
An example single value result is shown for the maximum temperature within the monoblock in Fig.~\ref{fig:probabilistic_max_temperature}.
For field data, each sample generates a full result field, but the variability of each point in that field is described by the same CDF ensemble structure as the single value results.

We perform the input sampling and surrogate execution using a combination of the SmartUQ\textsuperscript{\textregistered} software package and in-house python software.

\section{Credible model validation} \label{sec:Validation}

The purpose of model validation is to demonstrate the degree to which the model represents the true real world behaviour \citep{american2006guide}.
Critically, this is not a comparison of raw experiment and simulation results, because this would fail to account for the presence of uncertainty in \emph{both} datasets.
We select a validation metric by which to compute the model form uncertainty, a value that can then be added to a simulation prediction as additional epistemic uncertainty, such that there is a high degree of confidence that the true result lies within the simulation probability box.
Our use of probabilistic simulation based on thorough quantification of experiment uncertainty is what gives the above assertion credibility.
In this section we present our validation metric implementation, the comparison of experiment and simulation results, and the validation results.

\subsection{Implementation of a validation metric} \label{subsec:ValidationMetric}

A validation metric quantifies the model form uncertainty, that is the discrepancy in simulation results that cannot be mitigated by the uncertainty in either dataset \citep{ferson2008model,ferson2009validation,whiting2023assessment}.
We select the Modified Area Validation Metric (MAVM) \citep{voyles2014evaluation,voyles2015evaluation}, which computes model form uncertainty based on the area between simulation and experiment CDFs.
It does so for under- and over-prediction of the model separately, such that it captures asymmetry in the model form uncertainty.

\begin{figure}
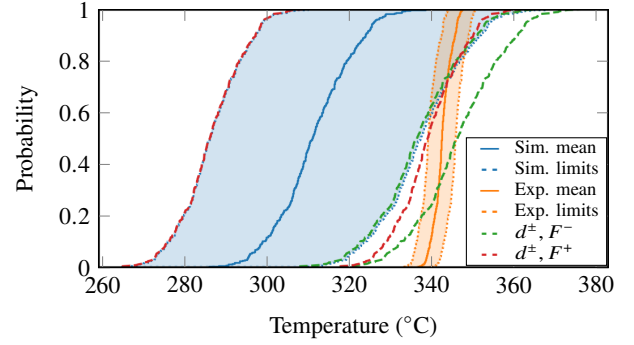

    \centering
    \inputtikz{mavm_implementation}
    \caption{An example of the implementation of the modified area validation metric (MAVM) for data with combined aleatoric and epistemic uncertainty. The MAVM derived under- and over-prediction of the model is computed for lower ($d^\pm, F^-$) and upper ($d^\pm, F^+$) bounds of the simulation probability box.}
    \label{fig:mavm_implementation}
\end{figure}

The MAVM computes two areas between experimental and simulation CDFs, the area for which the upper bound of the experiment probability box has a larger system response than the simulation CDF, $d^+$, and that for which the lower experimental bound has smaller system response, $d^-$.
These are defined as
\begin{align}
    d^+ &= \int_{-\infty}^{\infty} max \left(S_n^+(Y) - F(Y),0\right)dY \\
    d^- &= \int_{-\infty}^{\infty} max \left(F(Y) - S_n^-(Y),0\right)dY
\end{align}
where $Y$ is the system response, $F(Y)$ is a given simulation CDF, and $S_n^+(Y)$ and $S_n^-(Y)$ are the upper and lower bounding CDFs of the experiment probability box \citep{whiting2023assessment}.
What the MAVM results state is that the true predicted value lies within the range $[F(Y) - d^-, F(Y)+d^+]$.
In our validation, the probabilistic simulation also includes epistemic uncertainty, which is not directly accounted for in the definition of the MAVM.
We therefore extend the MAVM implementation to compute the MAVM results separately for each bounding CDF of the simulation probability box with respect to the bounding CDFs for the experimental data, as shown in Fig.~\ref{fig:mavm_implementation}.
The final validation result gives the range of possible simulation results as $\left[min([F(Y) - d^-]^i), max([F(Y)+d^+]^i)\right]$, where $^i$ is either the lower ($^-$) or upper ($^+$) simulation bound.
This provides the most conservative assessment of model form uncertainty given all combinations of simulation and experiment uncertainty.

We also require quantified validation of strain field prediction, utilising the high fidelity DIC data we have obtained from our experiments.
We apply our extended implementation of the MAVM to our field predictions on a point-wise basis, treating each point in the field as an independent result to be validated.
This provides a conservative assessment of the model form uncertainty over the field, constructed from the point-wise validation results.

We have implemented our validation metric methods as a module within the pyvale python package \citep{pyvale}.

\subsection{Validation results} \label{subsec:ValidationResults}
\begin{figure*}
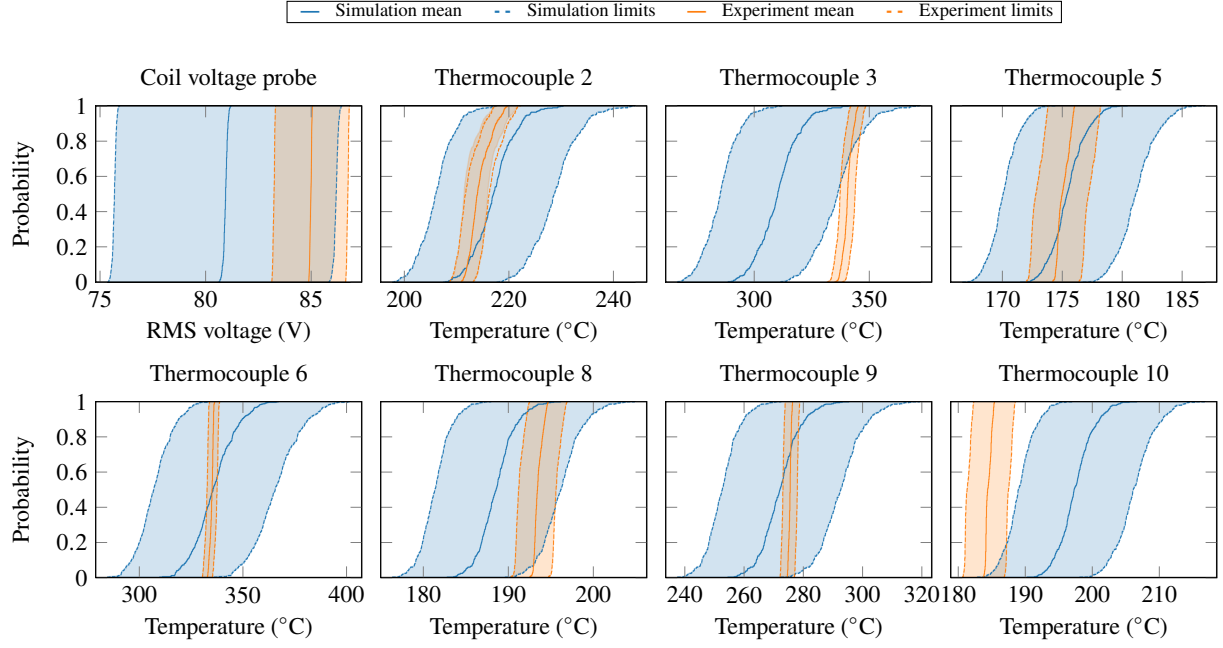

    \centering
    \inputtikz{results_point_comparison}
    \caption{Comparison of point sensor results between experiment and simulation, plotting full probability boxes for both.}
    \label{fig:results_point_comparison}
\end{figure*}

\begin{figure*}
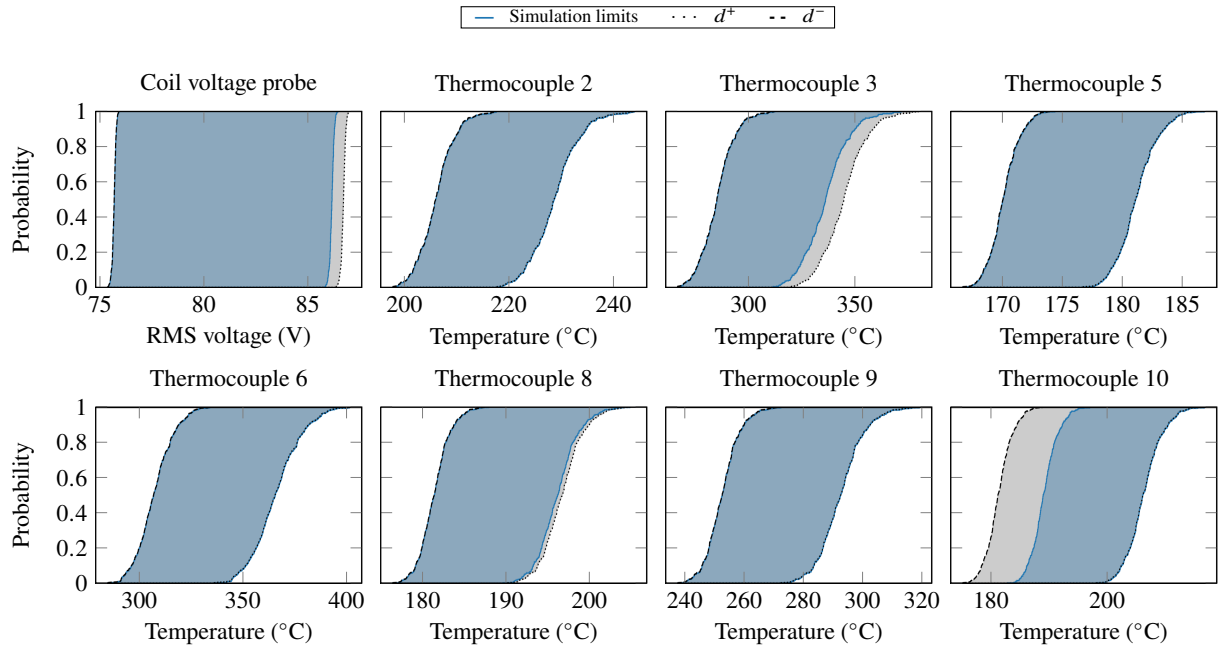

    \centering
    \inputtikz{results_point_validation}
    \caption{Simulation results with contribution from model form uncertainty, as derived using the validation metric.}
    \label{fig:results_point_validation}
\end{figure*}

Figure~\ref{fig:results_point_comparison} plots overlays of experiment and simulation probability boxes for the induction coil voltage probe and all functioning thermocouples.
This is followed by the corresponding validation metric results superposed on the simulation probability boxes in Fig.~\ref{fig:results_point_validation}.
When interpreting our validation metric there are two scenarios: 1) the ensemble of epistemic CDFs for the simulation completely overlaps that of the experiment, in which case the upper and lower validation metrics are zero, and 2) the ensemble of CDFs for the simulation and experiment contain some area which does not overlap and one or both of the upper and lower validation metrics are non-zero.
The first case is demonstrated in the results for thermocouples 2, 5, 6, and 9, where the experimental CDFs are completely contained within the bounds of the simulation CDFs and our validation metric upper and lower bounds are zero, collapsing onto the upper and lower simulation CDFs.
In this case, no model form uncertainty can be derived, and the predictive uncertainty is purely due to the experiment-derived uncertainty on model inputs.
The second case is demonstrated in the results for coil voltage, and thermocouples 3, 8, and 10.
Here, some of the experiment probability box lies outside that of the simulation, a discrepancy that is captured by the validation metric in the form of metric bounds that extend past the simulation probability box.
For these model outputs, the predictive uncertainty of the model due to input uncertainty must be augmented by model form uncertainty.
We note that the model form uncertainty for thermocouple 3 does not account for the discrepancy between the shape of the experiment and simulation CDFs.
This limitation of the MAVM is discussed in \S\ref{subsec:Improvment}.

\begin{figure*}
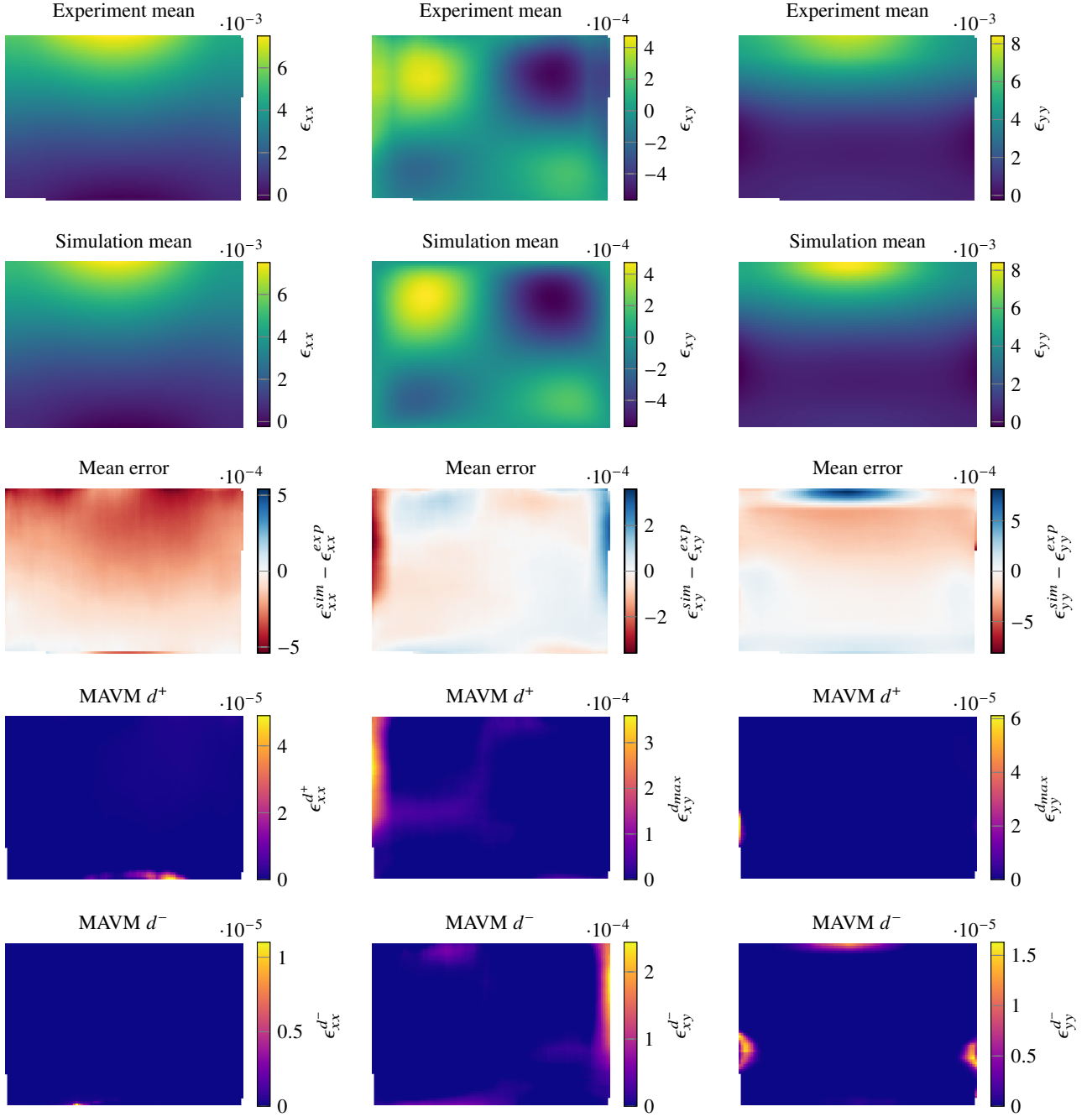

    \centering
    \inputtikz{results_field}
    \caption{Comparison of in-plane strain fields between experiment and simulation. Each plot set shows, from left to right, the mean experimental result, the mean simulation results, the mean error, and the validation metric result.}
    \label{fig:results_field}
\end{figure*}

Our strain field results are plotted in Fig.~\ref{fig:results_field}, showing the mean result from experiment and simulation, the mean error field, and the MAVM field.
These results demonstrate the value of both the data-rich diagnostic and our MAVM implementation in identifying the spatial distribution of model form uncertainty across the monoblock.
This allows a much more thorough understanding of the manner in which the model prediction does not align with the experimental data.
All strain field validation results highlight areas of the field in which the model is contributing uncertainty to the overall prediction uncertainty.
For the normal components of the strain tensor, $e_{xx}$ and $e_{yy}$, much of the mean error can be attributed to experiment uncertainty, though a small quantified model form uncertainty remains.
The shear component, $e_{xy}$, results demonstrate a different benefit of validation metric, as here the mean error results are not sufficiently conservative.
This would suggest that the probability distributions predicted for shear are not symmetric and require the rigour of the MAVM approach to identify the high discrepancy region of prediction space.

\section{Discussion} \label{sec:Discussion}

\subsection{Assessment of the initial test case} \label{subsec:AssessmentOfCase}

The validation results in \S\ref{subsec:ValidationResults} demonstrate that our finite element model is predicting the majority of outputs with little to no diagnosable model form error.
This is represented by little to no extension of the validation metric bounds beyond the probabilistic simulation probability box for those outputs.
The output with the most significant reported discrepancy is the shear component of the in-plane strain tensor.
Here the validation result suggests peak model form uncertainty of the order 100\% of the predicted result.
However, closer inspection of the experiment strain field and the MAVM plot in Fig.~\ref{fig:results_field} shows unphysical artefacts in the experimental strain field at the edge of the image that correspond with the location of maximum absolute MAVM values.
Visual comparison of the experimental and simulation fields suggests a possible spatial misalignment of the datasets, which would lead to the higher MAVM values around the inflection paths within the field.
This is plausible given the limitations of DIC to resolve up to the edge of samples and higher uncertainty in the data at those edges.
It is also possible that the uncertainty we assign to the relative position of the coil and monoblock is not sufficiently conservative.
While we can therefore explain our model discrepancy for this output, it does highlight a requirement for improved sensor uncertainty characterisation in our experiments.
In this instance it would be appropriate to only perform direct validation using an interior region of interest, thereby avoiding penalisation of the MAVM results due to DIC error in the edge regions.

\begin{figure}
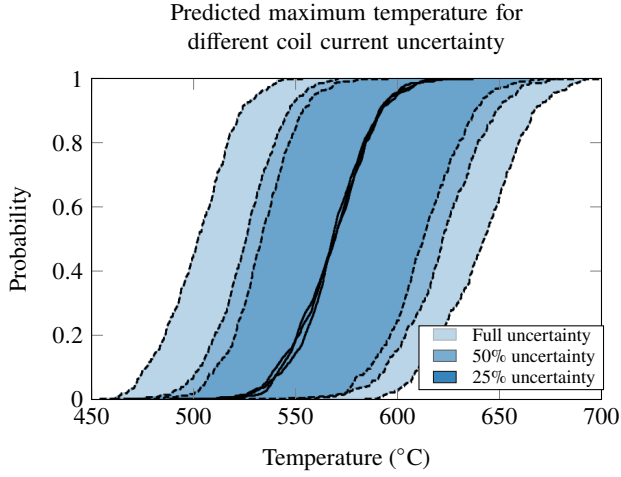

    \centering
    \inputtikz{current_uncertainty_reduction}
    \caption{The probabilistic maximum temperature result, shown for as reported current sensor epistemic uncertainty, and at 50\% and 25\% of the uncertainty range simulating improved sensor accuracy.}
    \label{fig:current_uncertainty_reduction}
\end{figure}

A notable feature of our simulation results is very wide epistemic uncertainty.
Our sensitivity analysis results in Fig.~\ref{fig:sensitivity} demonstrate the high sensitivity of all aspects of the system response to the coil current.
We note from Table \ref{tab:input_uncertainties} that the epistemic uncertainty on coil current is very high, which we can demonstrate contributes to a significant proportion of the overall epistemic uncertainty at output.
Figure~\ref{fig:current_uncertainty_reduction} plots the predicted maximum temperature within the monoblock for three coil current input scenarios: the as-reported uncertainty range, and 50\% and 25\% of this range.
The cause of this uncertainty is the accuracy of the Rogowski coil diagnostic for our particular experimental setup.
The sensor itself is capable of improved accuracy, but the geometry and proximity of the electrical feedthroughs in HIVE requires a more conservative assessment of accuracy.
With any validation experiment we can only assess model form uncertainty within the constraints of experimental uncertainty.
The large experimental uncertainty in our test case would therefore have to be reflected in predictive use of the model.
This does however highlight an additional benefit of our approach, in identifying the key source of uncertainty in our experiment.
If the validation the experiment provides is considered too uncertainty dominated, we are able to direct a decision maker to the most impactful improvement to the experiment.
This is discussed further in \S\ref{subsec:TowardsVirtualQualification}.

\subsection{Scope for methodology improvement} \label{subsec:Improvment}

While the methodologies we have presented are a significant improvement on traditional deterministic design-by-code approaches used in fusion, there are two specific areas we have identified that would merit specific work to improve the credibility of the validation approach.

The first is the epistemic sampling within the probabilistic simulation.
The requirement on this function is to compute the widest possible bounds of the epistemic uncertainty of all system \emph{outputs} given the defined variability of the inputs.
This is essentially an optimisation problem, and like all optimisation problems in a complex response space poses the problem of avoiding local optima and finding the global optimum.
The sampling approach we take is a simple, manual implementation of such an optimisation, increasing the density of uniform epistemic sampling until diminishing returns are observed.
We have identified two likely improvements to this approach we have taken: quantification of convergence of the results to a maximum output uncertainty band width, and implementation of more sophisticated sampling.
The former is likely to be a requirement for a probabilistic simulation to be considered credible, and also acts as a facilitator of the latter improvement.
The outer loop of the sampling, the epistemic sampling, would suit optimisation methods such as genetic algorithm \citep{mitchell1998introduction} or particle swarm \citep{kennedy1995particle} optimisation, in which the method is initially seeded across the whole input space.
This ensures as good a coverage as our current uniform sampling, and allows parallel searching of the input space to quickly identify regions of interest.
This will still result in increased compute requirements compared to our current approach, but would ensure efficient use of that resource.

The second area we wish to discuss is the ability of our validation metric to fully describe the model form uncertainty, specifically biased discrepancy.
This is best illustrated for thermocouples 3 and 10 in our results in Figures \ref{fig:results_point_comparison} and \ref{fig:results_point_validation}.
The overlay of the raw datasets for thermocouple 3 shows a higher likelihood of temperatures above $330~^\circ C$ in the experiment.
The validation metric records the underprediction, but the epistemic uncertainty it adds as model form uncertainty does not fully cover the observed discrepancy.
This is because the shape of the bounding CDF when including MAVM-computed model form uncertainty can only have the shape of either of the existing bounds of the probabilistic simulation probability box.
This suggests that there would be benefit in investigation alternative or improved validation metrics for use within our validation framework.

\subsection{Towards virtual qualification} \label{subsec:TowardsVirtualQualification}

We have presented this work in the context of a proposed framework for undertaking virtual qualification in fusion.
It is therefore important to consider how the methods and results we have presented would be used in a qualification scenario.
The benefit of the probabilistic approach to all components of the validation process is the quantification of probability of a component meeting a performance requirement, including an uncertainty band on this probability.
In realisation of the first fusion devices, decision makers will have to take on risk.
The probabilistic quantification of performance allows for more informed decision-making, with well understood trade off of performance and cost.
This type of approach is used in system design and safety assessment other industries \citep{safie2016role,pence2018methodology}, but not in fusion.
There is particular value of this approach in fusion, where over-conservative design risks the commercial viability of fusion as an energy source \citep{afify2024safety,elbez2024recommendations}.

A key benefit of our approach is the facilitation of cost-benefit analysis when assessing predicted component performance.
The outputs presented in Figures \ref{fig:results_point_validation} and \ref{fig:results_field} quantify the contribution to the uncertainty associated with a component meeting its requirements.
If we imagine these represent an operational component with predictions accounting for model form uncertainty, we can consider a requirement of temperature below some limit at each thermocouple, $T_{max} \le T_{lim}$.
Our approach generates a probability box describing a prediction of $T_{max}$ that accounts for input uncertainty and model form uncertainty.
The probability box is bounded by CDFs of $T_{max}$, $F^-$ and $F^+$.
We are therefore able to compute a range for the probability of the component meeting the requirement, $[P^-, P^+]$, where
\begin{align}
    P^- &= F^-\left(T_{lim}\right)\\
    P^+ &= F^+\left(T_{lim}\right).
\end{align}
A decision maker can either require amended design of the component to shift the whole probability box left, require better understanding of the epistemic uncertainty in the input space to reduce the width of the probability box, or require improvements to the model to remove or reduce the model form uncertainty.
If the cost of implementing each solution is known, our results allow an effective cost-benefit analysis to determine the best value approach.
This ability is further improved by the sensitivity results in Fig.~\ref{fig:sensitivity}, which identify the inputs for which reduced input uncertainty would be most beneficial.
The impact of various reductions in input uncertainty can be predicted using the probabilistic model, as has been done in Fig.~\ref{fig:current_uncertainty_reduction}, providing even more information to the cost-benefit analysis.
The relevance to fusion is emphasised when considering the scale, and therefore cost, of validation experiments and component testing in fusion relevant environments, such as those envisaged in the CHIMERA \citep{barrett2023chimera} and LIBRTI \citep{lawlessoverview} facilities.

\subsection{Future work} \label{subsec:FutureWork}   
This nature of this work is inherently forward-looking, and is intended to seed multiple avenues of future work.
High level, long term plans are described in Lawless et al. \cite{lawlessoverview} and presented in further detail in \S\ref{subsec:QualificationFusion}.
These plans culminate in how model validation can be \emph{assimilated} to operational environments with conditions that cannot be physically tested.
Before this can be attempted, our validation methods must be demonstrated on more complex systems, including those where experimental validation can only be conducted on parts of the simulation model in isolation.
This will require propagation of model form uncertainty through a simulation workflow, and characterisation of interactions of different physical phenomena such that model form uncertainty can be adjusted to account for this.
In all of the above, the usability of a model in physical component design relies on credible quantification of model form uncertainty for any output from the model, not just those directly validated.
The foundation of all of the above is credible model validation.
We believe that the work presented in this paper provides a detailed benchmark for its application to fusion problems, and fulfils the requirements of individual model validation in the virtual qualification framework.

\section{Conclusions} \label{sec:Conclusions}
Realisation of commercially viability fusion energy will depend on simulation-led qualification and on the ability of the fusion industry to make risk-based design decisions.
This will require a move away from the deterministic simulation and conservative design codes on which the industry has historically relied.
We have therefore presented an industrial application of rigorous VVUQ to a fusion-relevant test case.
This formalises an approach to credible model validation within a virtual qualification framework for fusion components and systems.
The main outcomes of this work are as follows:
\begin{enumerate}
    \item Development of a test case on which to demonstrate our probabilistic model validation framework that captures the complexities of using simulation in fusion component design. This includes tightly coupled multi-physics behaviour and uses a component synonymous with high heat flux heat sinks in fusion.
    \item Implementation of data-rich validation experiments on a fusion component, utilising a mixture of point sensors and image-based sensors. This work has highlighted the value of data-rich experiments, not only to provide high fidelity characterisation of system outputs, but also to directly measure all boundary conditions to which our model is sensitive.
    \item Training and validation of a hybrid Gaussian Process - modal decomposition surrogate model to facilitate rapid prediction of both point and field results, while quantifying the additional uncertainty this introduces.
    \item Sampling and propagation of experiment-derived probability boxes, incorporating aleatoric and epistemic uncertainty, to generate probability boxes describing predicted system behaviour. This included prediction of field responses.
    \item Novel implementation of the Modified Area Validation Metric, accounting for aleatoric and epistemic uncertainty in both experiment and simulation results and implementing on a point-wise basis to validate field predictions. We were therefore able to robustly identify the model form uncertainty of our finite element model.
    \item General agreement of the simulation with the experiment, to within the limits of the experimental uncertainty.
    \item Demonstration of the value of our approach in facilitating risk-aware decision making that maximises the cost-benefit of future work.
\end{enumerate}

We present this test case and all associated data as a benchmark for implementation of probabilistic validation in fusion and for development of model validation methodologies.

We conclude that the methods we have developed and demonstrated are credible, and believe that they can form the foundation for a virtual qualification framework for fusion.

\section*{CRediT authorship contribution statement}
\textbf{J. T. Horne-Jones}: Conceptualization, Data curation, Formal analysis, Methodology, Project administration, Software, Supervision, Validation, Visualization, Writing – original draft, Writing – review \& editing. \textbf{M. Baxter}: Conceptualization, Funding acquisition, Project administration, Resources, Supervision, Writing – review \& editing. \textbf{A. Tayeb}: Data curation, Formal analysis, Investigation, Writing – original draft. \textbf{L. Fletcher}: Conceptualization, Formal analysis, Methodology, Software, Validation, Visualization, Writing – review \& editing. \textbf{J. Paterson}: Investigation, Writing – review \& editing. \textbf{S. Biggs-Fox}: Formal analysis, Writing – original draft. \textbf{A. Harte}: Conceptualization, Funding acquisition, Project administration, Supervision.

\section*{Acknowledgment} 

We would like to thank the team who run the HIVE facility at UKAEA. We also thank the EPSRC Energy Programme for funding this work.

\section*{Funding Data}

This work has been funded by the EPSRC Energy Programme [grant number EP/W006839/1]. Dr Lloyd Fletcher and Dr Adel Tayeb acknowledge support from UKRI through the Future Leaders Fellowship scheme [grant number MR/Y015916/1]. To obtain further information on the data and models underlying this paper please contact PublicationsManager@ukaea.uk*. The data underlying this paper can be found at \url{https://doi.org/10.14468/7zzp-ny48}.


\if\prePrint0
    \begin{nomenclature}

    \entry{CDF}{cumulative distribution function}
    \entry{DAQ}{data acquisition system}
    \entry{DIC}{digital image correlation}
    \entry{HIVE}{Heating by Induction to Verify Extremes}
    \entry{ICP}{iterative closest point}
    \entry{MAVM}{modified area validation metric}
    \entry{PFC}{plasma facing component}
    \entry{TCx}{thermocouple x}
    \entry{UKAEA}{UK Atomic Energy Authority}

    \EntryHeading{Dimensionless Groups}

    \entry{Re}{Reynolds number}
    \entry{Pr}{Prandtl number}
    \entry{Nu}{Nusselt number}

    \end{nomenclature}
\else
    \section*{Nomenclature}

    \begin{tabular}{ll}
        CDF & cumulative distribution function \\
        DAQ & data acquisition system \\
        DIC & digital image correlation \\
        HIVE & Heating by Induction to Verify Extremes \\
        ICP & iterative closest point \\
        MAVM & modified area validation metric \\
        PFC & plasma facing component \\
        TCx & thermocouple x \\
        UKAEA & UK Atomic Energy Authority \\

        Re & Reynolds number \\
        Pr & Prandtl number \\
        Nu & Nusselt number \\
    \end{tabular}
\fi


\nocite{*} 

\bibliographystyle{asmejour}   

\bibliography{References} 




\appendix   

\section{DIC calibration, processing parameters and noise floors} \label{ap:dicCalibration}

\begin{table} [h!]
    \centering
    \begin{tabular}{|c|c|}
        \hline
        \textbf{Parameter}   & \textbf{Value} \\
        \hline
        $F_x$ Camera 0 [pixels] & 19447 \\
        $F_y$ Camera 0 [pixels] & 19449 \\
        $F_s$ Camera 0 [pixels] & 4 \\
        $\kappa_1$ Camera 0 & -0.1983 \\
        $\kappa_2$ Camera 0 & -1.1775 \\
        $\kappa_3$ Camera 0 & 25.011 \\
        $C_x$ Camera 0 [pixels] & 2622 \\
        $C_y$ Camera 0 [pixels] & 2292 \\
        $F_x$ Camera 1 [pixels] & 19488 \\
        $F_y$ Camera 1 [pixels] & 19487\\
        $F_s$ Camera 1 [pixels] & -3\\
        $\kappa_1$ Camera 1 & -0.1868\\
        $\kappa_2$ Camera 1 & -1.1657\\
        $\kappa_3$ Camera 1 & 25.756\\
        $C_x$ Camera 1 [pixels] & 2656\\
        $C_y$ Camera 1 [pixels] & 2255\\
        $T_x$ [mm] & -76.16\\
        $T_y$ [mm] & 0.37\\
        $T_z$ [mm] & 6.68\\
        $\theta$ [$^{\circ}$] & -0.18\\
        $\phi$ [$^{\circ}$] & 11.22\\
        $\psi$ [$^{\circ}$] & -0.27\\
        \hline
    \end{tabular}
    \caption{Stereo DIC calibration parameters.}
    \label{tab:CalParam}
\end{table}

\begin{table}
    \centering
    \begin{tabular}{ |c|p{2.5cm}|p{2.5cm}| } 
        \hline
        \textbf{Quantity} & \textbf{Average temporal noise floor} & \textbf{Maximum temporal noise floor} \\
        \hline
        \textbf{$E_{xx}$ ($\mu \varepsilon$)} & 25.6 & 116.8\\
        \hline 
        \textbf{$E_{yy}$ ($\mu \varepsilon$)} & 23.2 & 76.2 \\
        \hline
        \textbf{$E_{xy}$ ($\mu \varepsilon$)} & 11.2 & 81.8\\
        \hline
    \end{tabular} 
    \caption{DIC resolution from temporal noise floor analysis.}
    \label{tab:NoiseFloor}
\end{table}

\FloatBarrier

\section{Thermocouple placement} \label{ap:ThermocouplePlacement}

\begin{table} [h!]
    \centering
    \begin{tabular}{ |c|c|c|c| } 
        \hline
        \textbf{Thermocouple} & \textbf{x (m)} & \textbf{y (m)} & \textbf{z (m)} \\
        \hline
        TC1 & 0.0116 & -0.0245 & 0.0194 \\
        TC2 & 0.0138 & -0.0245 & 0.0013 \\
        TC3 & 0.0067 & -0.0245 & 0.0124 \\
        TC4 & 0.0110 & 0.0245 & 0.0031 \\
        TC5 & -0.0105 & 0.0245 & -0.005 \\
        TC6 & -0.0058 & 0.0245 & 0.0171 \\
        TC7 & -0.018 & -0.0006 & 0.0164 \\
        TC8 & -0.018 & -0.004 & -0.0085 \\
        TC9 & -0.018 & -0.0047 & 0.0073 \\
        TC10 & -0.018 & 0.0124 & -0.0032 \\
        \hline
    \end{tabular} 
    \caption{Coordinates used for machining of thermocouple installation holes. The coordinate system is located on the pipe axis, at the midpoint of the pipe length.}
    \label{tab:ThermocouplePlacement}
\end{table}

\FloatBarrier


\end{document}